\documentclass[twocolumn]{revtex4-2}
\usepackage{graphicx} 
\usepackage{amsmath}
\usepackage{amssymb}
\usepackage{multirow}
\usepackage{booktabs}  
\usepackage{appendix}
\usepackage{placeins}  
 
\usepackage{amsfonts}
\usepackage{xcolor}
\usepackage{enumerate}
\usepackage{soul}
\usepackage{bm}

\usepackage{caption}
\usepackage{subcaption}
\usepackage{multirow}
\usepackage{mhchem}
\usepackage{adjustbox}
\usepackage[normalem]{ulem}

\def\rr{{\bf r}}
\def\mm{{\bf m}}

\begin{document}

\title{Equivariant Many-body Message Passing  \\ 
    Interatomic Potentials 
    for Magnetic Materials}

\author{
Cheuk Hin Ho$^{1,2}$,
Cas van der Oord$^{2}$,
James P. Darby$^{2}$,
Theo Keane$^{3}$,
Raz L. Benson$^{3}$,
Cristian Rebolledo Espinoza$^{3}$,
Rutvij Kulkarni$^{3}$,
Elina Spinu$^{3}$,
Michail Papanikolaou$^{3}$,
Richard Tomsett$^{3}$,
Robert M. Forrest$^{3}$,
Jonathan J. Bean$^{3,*}$,
G\'abor Cs\'anyi$^{2,4,5,*}$,
Christoph Ortner$^{1,2,}$
}

\thanks{\texttt{christoph.ortner@ubc.ca}}
\thanks{\texttt{gc121@cam.ac.uk}}
\thanks{\texttt{jonathan@materialsnexus.com}}

\affiliation{
$^{1}$Department of Mathematics, University of British Columbia \\
$^{2}$Symmetric Group, 71-75 Shelton Street, Covent Garden, London, WC2H 9JQ, United Kingdom \\
$^{3}$MatNex, Salisbury House, Station Road, Cambridge, CB1 2LA, United Kingdom \\
$^{4}$Engineering Laboratory, University of Cambridge, Trumpington St, Cambridge, United Kingdom\\
$^{5}$Max Planck Institute for Polymer Research, Ackermannweg 10, Mainz, 55128, Germany
}

\date{\today}

\begin{abstract}
Magnetism governs key properties of materials used in energy, data storage, and spintronic technologies, yet its complex coupling to lattice and electronic degrees of freedom challenges conventional first-principles approaches. We introduce an equivariant message-passing graph neural network that embeds atomic magnetic moments as explicit degrees of freedom, enabling the learning of magnetic interactions beyond collinear approximations. The model learns physically consistent and transferable representations of magnetic behaviour and can incorporate spin–orbit coupling, achieving near density-functional-theory accuracy with strong data efficiency across diverse magnetic systems by fine-tuning from a pre-trained model. Applications to structural transformations, finite-temperature magnetic phenomena, and materials screening for strongly spin-orbit coupled materials demonstrate transferable magnetic behaviour, establishing a practical foundation for data-driven, high-throughput discovery of complex magnetic materials.
\end{abstract}
\maketitle

\section{Introduction}

Magnetism plays a fundamental role in determining the properties of a wide range of materials central to modern technology, including structural alloys, permanent magnets, and functional electronic and spintronic systems. These applications span information storage~\cite{ZuticSpintronics2004, sverdlov2025editorial}, turbine generators~\cite{pavel2017role, GutfleischMagMaterials2011}, electronic devices~\cite{cui2018current}, and robotic systems~\cite{campbell1994permanent, XuRobtoActuators2015}. They increasingly extend to the emerging study of two-dimensional materials, where magnetic effects enable qualitatively new physical behavior~\cite{aldea2025challenges}. Across these systems, magnetic degrees of freedom are not a perturbative correction but give rise to qualitatively distinct physical behavior.

Accurately describing such behavior requires a consistent treatment of both atomic structure and electronic degrees of freedom, including spin. In particular, the coupling between atomic positions and magnetic moments gives rise to a complex, high-dimensional energy landscape with multiple competing configurations. This makes predictive modeling of magnetic materials inherently challenging, especially when exploring different magnetic orderings or thermodynamic conditions.

Density functional theory (DFT) has been the primary computational tool for capturing magnetic effects, but its high computational cost limits simulations to small systems and short time scales, restricting the study of finite-temperature and mesoscale phenomena.
Recent advances in machine-learned interatomic potentials (MLIPs) have extended atomistic modeling to larger systems while retaining near-DFT accuracy.
Both descriptor-based \cite{rinaldi2024non, novikov2022magnetic, drautz2020atomic} and neural-network–based models \cite{yu2024spin} have been developed for magnetic systems. 
Yu et al.~\cite{yu2024spin} introduced Heisenberg-edge and spin-distance edge graph neural networks (SEGNNs), which incorporate magnetic interactions through the dot products of spin vectors.
This scalar treatment limits expressivity and cannot capture the full vectorial and orientation-dependent character of non-collinear magnetism.
Behler and co-workers~\cite{eckhoff2021high} proposed spin-dependent atom-centered symmetry functions within their ``high-dimensional neural network potentials''; however their architecture remains restricted to collinear spin configurations. Xu et al.~\cite{xu2025spin} proposed a spin-informed graph neural network that remains limited to collinear magnetism. Rinaldi et al.~\cite{rinaldi2024non} proposed a non-collinear linear model that incorporates magnetic interactions but excludes spin–orbit coupling, and requires assembling thousands of configurations for a single element while at the same time relying on constrained DFT, which can possibly amplify inaccuracies and produce training databases that are challenging to fit~\cite{rinaldi2024non}. Hu et al.~\cite{zheng2025integrating, yu2024spin} developed the DeepSPIN framework, which uses a ``pseudo-atom representation'' to encode the coupling between lattice and spin configurations as interdependent features.

In this work, we present an equivariant message-passing graph neural network architecture (mMACE) that introduces atomic magnetic moments as explicit equivariant degrees of freedom. 
Our implementation builds on the MACE architecture~\cite{batatia2022mace}, but the modifications we introduce are generic and applicable to other frameworks.
The magnetic moments transform under the symmetry group appropriate to the physics—O(3) when spin–orbit coupling is present (or, O(3)$\times$O(3) when it is not) yielding a systematically improvable description of non-collinear magnetism and spin–orbit effects that includes collinear magnetism as a special case, all at modest additional computational cost.

We demonstrate the accuracy and broad applicability of mMACE across a range of magnetic phenomena:
On collinear benchmarks (FeAl, CrN), mMACE reduces force and stress errors by factors of three to five relative to existing magnetic machine-learned potentials (c.f. Section~\ref{sec:public_datasets}).
When trained on large foundational datasets (MATPES, MP-ALOE), the model systematically improves on magnetically active configurations while maintaining accuracy elsewhere, and serves as an effective starting point for targeted fine-tuning with minimal additional data (c.f. Section~\ref{sec:pretrained_model_training}): a small number of configurations suffices to recover the FeNi Bain-path energetics across magnetic states (c.f. Section~\ref{sec:FeNi binary system}), and the frustrated non-collinear ground state of Mn$_3$Pt is reliably found from random initial spin orientations that are outside the training dataset (c.f. Section~\ref{sec:Mn3Pt}).
Beyond energies and forces, the framework enables prediction of derived magnetic properties: spin–orbit coupling is incorporated to resolve magnetocrystalline anisotropy at the sub-meV scale across chemically diverse structures (c.f. Section~\ref{sec:magnetocrystalline_anisotropy}), Heisenberg exchange constants are extracted in quantitative agreement with first-principles references, and mMACE-driven Monte Carlo simulations yield Curie temperatures in closer agreement with experiment than classical Heisenberg models~(c.f. Section~\ref{sec:non_collinear_fe}).

\section{Methods}
\label{sec:theory}
\subsection{Incorporating magnetism into EMPNNs}
Equivariant message passing neural networks (EMPNNs) have been widely used to design MLIPs as surrogate models for DFT \cite{batatia2022mace, batzner20223, bochkarev2024graph, gasteiger2021gemnet, haghighatlari2022newtonnet, satorras2021n}. We outline a framework for EMPNN parameterization of potential energy surfaces that include atomic magnetic moments $\mathbf{m}_i$ as atomic features. They can be thought of as a coarse-grained projection of the spin density onto atomic sites, either via a specified partitioning scheme or latent variables that are implicitly defined during model training. 

An atomic structure is described by a graph with nodes indexed by $i$ representing atoms and edges $(i, j)$ specifying an atomic neighbourhood relation $j \in \mathcal{N}(i)$. In each layer $t$ of an EMPNN, the \textit{state} of a node (atom) $i$ is described by a tuple
\begin{equation}
\label{eqn:general_node_state}
    \sigma^{(t)}_i := (\mathbf{r}_i, \mathbf{m}_i, z_i, h^{(t)}_i),
\end{equation}
where $\mathbf{r}_i \in \mathbb{R}^3$ denotes the atomic position, $\mathbf{m}_i \in \mathbb{R}^3$ the (generally non-collinear) atomic magnetic moment, $z_i \in \mathbb{Z}$ the atomic number (chemical element) and $h^{(t)}_i$ the hidden features at layer $t$, which can be scalar, vector or higher order. The atomic neighbourhood, denoted by $\mathcal{N}(i)$, is the set of all atoms $j$ within a cutoff radius $r_{\rm cut}$ from atom $i$.
At each message passing step, a message $\mathcal{M}^{(t)}_i$ is constructed by aggregating information from neighbouring atoms
\begin{equation}
    \mathcal{M}^{(t)}_i := \bigoplus_{j \in \mathcal{N}(i)} M_t(\sigma^{(t)}_i, \sigma^{(t)}_j)
\end{equation}
where $M_t$ is a learnable message passing function and $\bigoplus_{j \in \mathcal{N}(i)}$ is a learnable permutation invariant pooling operation. At each update step, the message $\mathcal{M}^{(t)}_i$ is then transformed into a new feature vector for the next layer through a learnable update function $U_t$
\begin{equation}
    h^{(t+1)}_i := U_t(\sigma^{(t)}_i, \mathcal{M}^{(t)}_i).
\end{equation}
After $t_{\rm max}$ layers of message passing and updates (in our setting, $t_{\rm max}=2$ is often sufficient), learnable readout functions $\mathcal{R}_t$ map the node states to interaction site energies (atomic energies)
\begin{equation}
    E_i^{\rm inter} := \sum^{t_{\rm max}}_{t=1} \mathcal{R}_t(h^{(t)}_i).
\end{equation}
Finally, the total potential energy is obtained by summing the contributions from all sites,
\begin{equation}
    E := \sum_{i} E_i^{\rm inter} + \sum_i E_{0,z_i} (|\mathbf{m}_i|)
\end{equation}
and one-body contributions $E_0$ that depend only on the atomic magnetic moments. Forces are obtained by taking the negative gradients with respect to atomic positions, 
\begin{equation}
\label{eqn:usual_forces}
    F_{i} := -\nabla_{\mathbf{r}_i} E\left(\{ \mathbf{r}_i, \mathbf{m}_i, z_i)\}_i \right).
\end{equation}
The corresponding magnetic forces on the atomic magnetic moment of atom $i$ are given by
\begin{equation}
    F^{\rm mag}_{i} := -\nabla_{\mathbf{m}_i}  E\left(\{ \mathbf{r}_i, \mathbf{m}_i, z_i)\}_i \right).
\end{equation}
These gradients enable deterministic, gradient-based optimization of the magnetic degrees of freedom to obtain the equilibrated (self-consistent) total energy. For periodic systems, stresses (or virials) on the unit cell can be derived in an analogous fashion to typical MLIPs.

Evaluating the energy in a configuration $\{{\bf r}_i, {\bf m}_i, z_i\}_i$ with non-zero magnetic forces $\nabla_{\mathbf{m}_i}E \neq 0$ effectively corresponds to a constrained DFT calculation, in which the magnetic moments are held fixed at non-equilibrium orientations or magnitudes. In practice, effective forces acting on atoms may be defined in different ways depending on the intended application. For example, in equilibrium geometry optimization, one typically relaxes the magnetic degrees of freedom via self-consistent cycles (at fixed atomic positions) and evaluates nuclear forces on the relaxed surface. In contrast, for spin-lattice dynamics, one uses the instantaneous forces evaluated at the evolving spin configuration.

\begin{figure*}[t]
    \centering
    \begin{minipage}{\textwidth}
        \centering        \includegraphics[width=\textwidth]{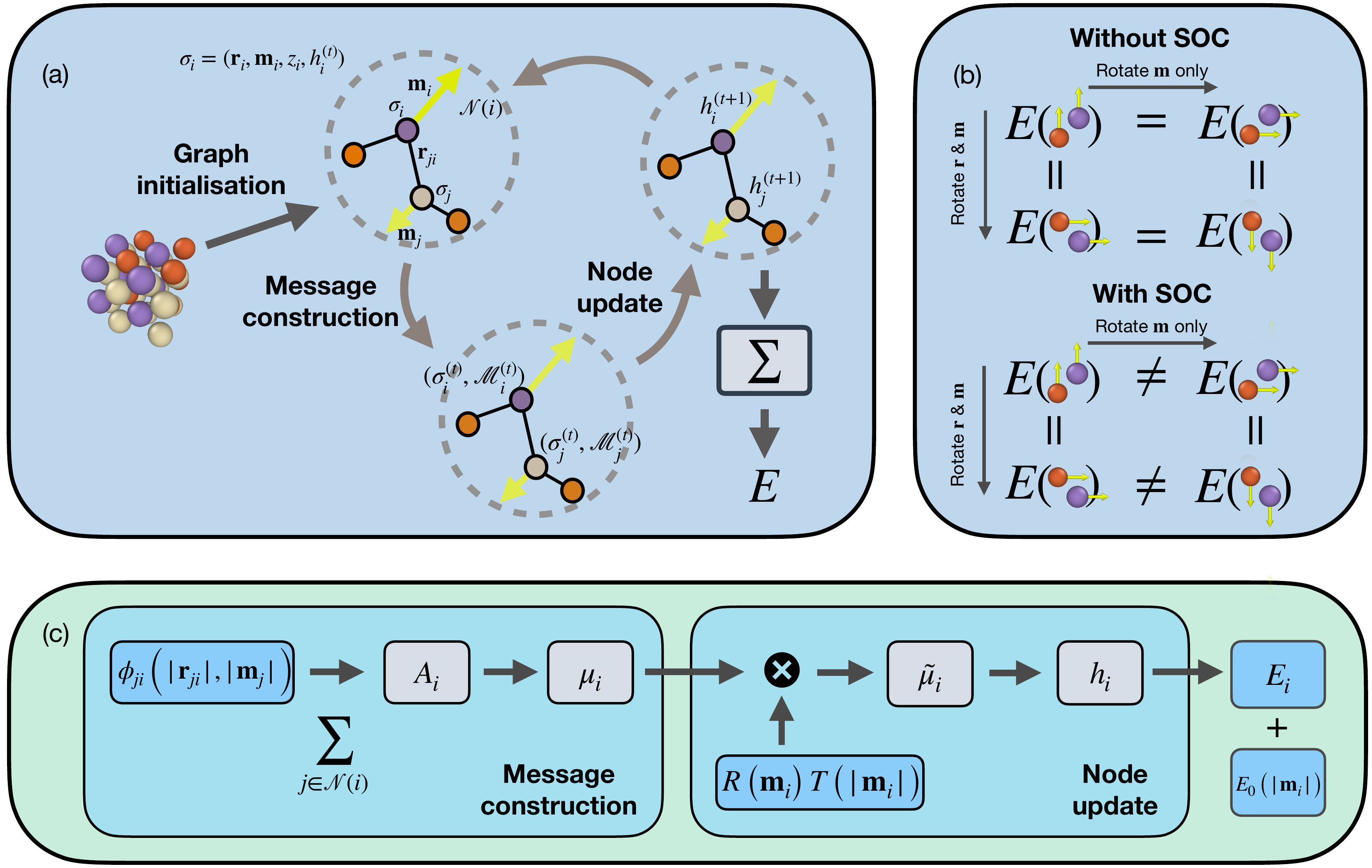}
    \end{minipage}
    \vspace{5pt}
    \caption{Schematic overview of the magnetic message-passing architecture in mMACE.
(a) Graph construction and message-passing process, where each node represents an atom with associated position, species, and magnetic moment. Messages exchanged along atomic edges encode magnetic moment-dependent interactions and are aggregated to produce local atomic energies (b) Inclusion of spin–orbit coupling (SOC) breaks rotational degeneracy, enabling the model to capture magnetocrystalline anisotropy.
(c) Neural network architecture of mMACE, consisting of coupled positional and magnetic message-construction blocks and node-update operations. The network outputs atomic energies, forces, magnetic forces, and stress tensors while maintaining full rotational equivariance.}
\label{fig:general_framework_combined}
\end{figure*}

\subsection{Equivariance and spin-orbit coupling}
$O(3)$-equivariant MPNNs utilize architectures for $M_t, U_t, \mathcal{R}_t$ such that the hidden features $h_i^{(t)}$ and outputs $E_i$ transform under rotation in a precise, consistent way. In most MPNN based MLIPs, a rotation $Q$ acts only on the positions ${\bf r}_i \mapsto Q {\bf r}_i$. Introducing magnetic degrees of freedom means that rotations can now also act on the atomic magnetic moments ${\bf m}_i$. Therefore we need to determine the  symmetry relations that the outputs $E_i$ should satisfy under various operations involving these new degrees of freedom. 

The total energy of the system must always remain invariant under a rotations that act simultaneously on both atomic positions and magnetic moments, \textit{i.e.},
\begin{equation}
    \label{eq:generic_rot_inv}
    \begin{split}
    E\big( \{ Q\rr_i, Q\mm_i, z_i \}_i \big) 
	=
    E\big( \{ \rr_i, \mm_i, z_i \}_i \big) & \\ 
    \forall Q \in O(3). & 
    \end{split}
\end{equation}
The situation becomes more subtle when considering the action of two independent rotations on positions $\mathbf{r}_i$ and moments $\mathbf{m}_i$, \textit{i.e.} whether we have the more restrictive symmetry 
\begin{equation}
    \label{eq:nospinorbitcoup}
    \begin{split}
    E\big( \{ Q \rr_i, Q' \mm_i, z_i \}_i \big) 
=
    E\big( \{ \rr_i, \mm_i, z_i \}_i \big) & \\ 
    \forall Q, Q' \in O(3), &
    \end{split}
\end{equation}
that is, invariance under the product group $O(3) \times O(3)$. 

Physically, imposing \eqref{eq:nospinorbitcoup} means that the orientation of magnetic moments to the atomic geometry are decoupled, that is, the system does not exhibit spin–orbit coupling (SOC). Conversely, if we only impose \eqref{eq:generic_rot_inv} (but do not require \eqref{eq:nospinorbitcoup}) then we break the independent rotational symmetries of position and magnetic space. This symmetry breaking is required if the system does exhibit SOC.

This decision leads to different group symmetries and hence different choices in the design of the EMPNN model architecture. An architecture {\em with} SOC can be obtained as a relatively straightforward extension of existing $O(3)$-equivariant MPNN architectures, whereas an architecture {\em without} SOC requires a more careful extension to account for the product group $O(3) \times O(3)$. In the next section we explain the architecture for the SOC case and include the extension to a non-SOC architecture in Appendix~\ref{app:sec:non_soc_model}.

\subsection{Model architecture}
Building on the foregoing symmetry considerations, we introduce the spin-coupled magnetic MACE (mMACE) architecture, which is equivariant under joint rotations of positions and magnetic moments, \textit{i.e.}, satisfying \eqref{eq:generic_rot_inv}. This choice maintains the flexibility to capture SOC effects while keeping the architecture a straightforward extension of the standard MACE framework. When training on data without SOC, the more stringent symmetry \eqref{eq:nospinorbitcoup} can be effectively achieved to within high precision through data augmentation; further details are provided in Appendix~\ref{app:sec:data_augmentation}. An alternative architecture that explicitly enforces the stronger $O(3) \times O(3)$ symmetry \eqref{eq:nospinorbitcoup} is presented in Appendix~\ref{app:sec:non_soc_model}. 

Let $\sigma^{\rm init}_i := (\mathbf{r}_i, \mathbf{m}_i, z_i) \in \mathbb{R}^3 \times \mathbb{R}^3 \times \mathbb{Z}$ denote the initial state of an atom.
At each message passing iteration $t = 1, 2, \dots$, node features from the previous layer are denoted by $h^{(t)}_{i,k l_2 m_2}$. For the first layer ($t = 1$) we only allow invariant features (hence only the values $l_2 = m_2 = 0$ are required), which are initialized as a learnable embedding of the chemical elements, 
\begin{equation}
    h^{(1)}_{i,k00} := W_{k z_i}.
\end{equation}
This defines the node state $\sigma_i^{(1)} := (\mathbf{r}_i, \mathbf{m}_i, z_i, h^{(1)}_i)$ for message passing, as in \eqref{eqn:general_node_state}.

For each atom $i$, information along its edges $(i, j)$ is embedded using two learnable invariant bases $P_{k l_1 l_2 l_3}$ and $K_{k l_1 l_2 l_3}$
\begin{align}
    J_{ji} &:= J(|\mathbf{r}_{ji}|, z_i, z_j), & \\ 
    T_j &:= T(|\mathbf{m}_{j}|, z_j), \\
    \label{eqn:pos_radial}
    P^{(t)}_{ji, k l_1 l_2 l_3} &:= {\rm MLP}^{(t), {\rm P}}_{k l_1 l_2 l_3}\left( J_{ji} \oplus T_j \right), \\
    \label{eqn:mag_radial}
    K^{(t)}_{ji, k l_1 l_2 l_3} &:= {\rm MLP}^{(t), \rm K}_{k l_1 l_2 l_3}\left( J_{ji} \oplus T_j \right), 
\end{align}
where $\oplus$ denotes a concatenation, and $J$ and $T$ are (non-trainable) Bessel and Chebyshev polynomial functions with pre-defined degree and element-dependent transformation; details are provided in Appendix~\ref{app:sec:radial_basis}. We did not include $|\mathbf{m}_i|$ in the embedding of the edge; these will be included at a later stage in the architecture.  

Both MLPs acts over the batch dimension (nodes or edges),
and transform the concatenated input features to an output feature with dimension indexed by $k l_1 l_2 l_3$. The construction of $P$ and $K$  are analogous to the learnable radial basis $R$ in the original MACE architecture~\cite{batatia2022mace}, but now incorporate information about the spin magnitudes $|\mathbf{m}_j|$, which is an additional invariant input.

Next, the equivariant edge features with magnetic information can be constructed through the following sequence of operations
\begin{align}
\label{eqn:edge_embed_pos}
\phi^{(t),{\rm pos}}_{ji, k l_3 m_3}
&:= \sum_{l_1 m_1 l_2 m_2} 
    C^{l_3 m_3}_{l_1 m_1 l_2 m_2}  \\
    \notag
&\hspace{1.6cm} \cdot P^{(t)}_{ji, k l_1 l_2 l_3}  Y^{m_1}_{l_1}(\hat{\mathbf{r}}_{ji}) 
    h^{(t)}_{j, k l_2 m_2} \\
\label{eqn:edge_embed_mag}
 \phi^{(t),{\rm mag+pos}}_{ji, k l'_3 m'_3}
& 
:= \sum_{l'_1 m'_1 l'_2 m'_2} 
    C^{l'_3 m'_3}_{l'_1 m'_1 l'_2 m'_2} \\
    \notag 
&
\hspace{1.6cm} \cdot K^{(t)}_{ji, k l'_1 l'_2 l'_3} R^{m'_1}_{l'_1}(\mathbf{m}_{j}) 
    \phi^{(t),{\rm pos}}_{ji, k l'_2 m'_2}
, 
\end{align}
where $\hat{\mathbf{r}}_{ji} := \mathbf{r}_{ji} / |\mathbf{r}|$, $Y^{m_1}_{l_1}$ are real spherical harmonics, and $R^{m'_1}_{l'_1}$ are real solid harmonics which guarantee smoothness in the limit $|\mathbf{m}_j| \rightarrow 0$. When $t = 1$, \eqref{eqn:edge_embed_pos} simplifies to
\begin{equation}
\phi^{(1),{\rm pos}}_{ji, k l_3 m_3} := P^{(t)}_{ji, k}  Y^{m_3}_{l_3}(\hat{\mathbf{r}}_{ji}) h^{(1)}_{j, k 0 0},
\end{equation}
reducing the computational cost of the first layer.
The two contractions \eqref{eqn:edge_embed_pos} and \eqref{eqn:edge_embed_mag} can be merged into a single operation, requiring only one learnable radial basis, however, the resulting intermediate tensor would be significantly larger, leading to increased computational cost and memory usage. For this reason, we choose to break the contraction into two stages. 

Next, the two-body basis on each node can be formed by pooling over the neighbourhood of atom $i$
\begin{equation}
A^{(t)}_{i, k l m} := 
    \sum_{j \in \mathcal{N}(i)} 
    \phi^{(t),{\rm mag + pos}}_{ji, k l m}.
\end{equation}
Optionally, density normalization \cite{batatia2025foundation} can be applied to the two-body features to ensure internal normalization of the weights, which can result in better training dynamics and smooth extrapolation to high pressure.

Once the two-body basis $A^{(t)}_{i, klm}$ is formed, the many-body basis for each correlation order $\nu$ is 
\begin{align}
\mathbf{B}^{(t)}_{i, \eta_\nu k LM} & := \sum_{\mathbf{lm}} C_{\eta_\nu \mathbf{lm}}^{LM} \prod_{\xi=1}^{\nu} \sum_{\tilde{k}} W^{(t)}_{k \tilde{k}  l_\xi} A_{i,\tilde{k} l_\xi m_\xi}^{(t)},
\end{align}
where $C^{L M}_{\eta_\nu \mathbf{l m}}$ are generalized Clebsch–Gordan coefficients~\cite{batatia2023general}. Before we define the many-body message $\mathcal{M}^{(t)}$ of the current iteration $t$, we first specify intermediate features, which are linear in the many-body basis, 
\begin{equation}
    \mu^{(t)}_{i,kLM} := \sum_{\nu = 1}^{\nu_{\rm max}} \sum_{\eta_\nu} W_{z_i \eta_\nu kL}^{(t),\nu} \mathbf{B}^{(t)}_{i,\eta_\nu kLM}.
\end{equation}
We correlate the magnetic moment of the center atom $i$ itself with its atomic environment, resulting in the many-body message
\begin{align}
    Q^{(t)}_{i, k l_0 l_1 L} &
    := {\rm MLP}^{(t), {\rm Q}}_{k l_0 l_1 L}\left( T(|\mathbf{m}_{i}|, z_i) \right), \\
    \mathcal{M}^{(t)}_{i, kLM} &
    := \!\! \sum_{l_0 m_0 l_1 m_1} \!\! C^{LM}_{l_0 m_0 l_1 m_1} Q^{(t)}_{i, k l_0 l_1 L}R^{m_0}_{l_0}(\mathbf{m}_i) \mu^{(t)}_{i, k l_1 m_1}.
\end{align}
Here $Q^{(t)}_{i, k l_0 l_1 L}$ is again a learnable radial function analogous to \eqref{eqn:mag_radial} but depends only on the element and magnitude of magnetic moment of atom $i$.

The hidden node feature update is completed via a residual connection
\begin{align}
    h^{(t+1)}_{i, kLM}
    := & 
    {\sum_{\tilde{k}} W_{kL,\tilde{k}}^{(t),\mathrm{self}} \mathcal{M}_{i,\tilde{k}LM}^{(t)}}
    + \sum_{\tilde{k}} W_{kL,\tilde{k}}^{(t)} \mu_{i,\tilde{k}LM}^{(t)} \notag \\
    & + \sum_{\tilde{k}} W_{k z_i L,\tilde{k}}^{(t)} h_{i,\tilde{k}LM}^{(t)}.
\end{align}
The atomic {\em interaction energies} are then obtained as a readout applied to the invariant node features, 
\begin{align}
\label{eqn:one_body_magmom_inter}
    E_i^{\rm inter} 
    := \sum_{t = 1}^{t_{\rm max}} \left( \mathcal{R}_t\left(h^{(t)}_{i, \bar{k}LM}\right) \right). 
\end{align}
The readout functions \(\mathcal{R}_t\) are implemented as nonlinear MLPs in the final layer, while all preceding message-passing layers remain linear.

Finally, a one-body magnetic moment contribution is introduced to ensure the correct large volume limit (\textit{i.e.} as $|\mathbf{r}_{ji}| \rightarrow \infty$ for all $i,j$). This term is pre-calculated as the energy of isolated atoms in different magnetic states or fitted using a simple linear model, 
\begin{equation} 
    \label{eqn:one_body_magmom_1b}
    E_{0, z_i}(|\mathbf{m}_i|) := \sum_{n, z} W_{n z} \delta_{z z_i} \tilde{T}_n(|\mathbf{m}_i|), 
\end{equation}
where $(\tilde{T}_n)_n$ is a Chebyshev polynomial basis with an appropriate transformation applied to the input (See Appendix~\ref{app:sec:one_body_mag_contribution_mlips} and Appendix~\ref{app:sec:radial_basis} for details).

The final atomic energies are given by the sum of interaction and one-body contributions, 
\begin{align}
\label{eqn:one_body_magmom}
    E_i := E^{\rm inter}_i 
    {+ E_{0, z_i}(|\mathbf{m}_i|)}.
\end{align}
Further details on implementation can be found in Appendix~\ref{sec:implementation_details}. The implementation of mMACE for reproducing the experiments is available at \cite{ho2026magneticmace}. Details on parameter estimation (``training'') and dataset generation can be found in Appendix~\ref{app:sec:parameter_estimation} and \ref{app:sec:data_generation}.


\section{Results}

\subsection{Training from Scratch: Benchmarks on Public Datasets}
\label{sec:public_datasets}

\subsubsection{FeAl benchmark}

We first benchmark the mMACE architecture on the collinear spin-polarized FeAl dataset of Kotykhov \textit{et al.}~\cite{kotykhov2023constrained}, which was originally published for fitting a magnetic MTP (mMTP) model. Figure~\ref{fig:benchmark_with_tables}a compares energy, force and stress predictions of MACE and mMACE against the DFT reference. The standard MACE model exhibits regions where it predicts zero forces and stresses despite non-zero DFT values, reflecting its inability to capture magnetic moment contributions. In contrast, mMACE shows no such artefacts, indicating that the explicit inclusion of magnetic degrees of freedom enables it to represent spin-dependent energy surfaces more accurately.

These observations are quantified in Fig.~\ref{fig:benchmark_with_tables}a,c, where mMACE achieves the lowest RMSE across energies, forces, and stresses on the test set of \cite{kotykhov2023constrained}, outperforming both mMTP and the standard MACE model. In particular, the force RMSE is reduced by nearly a factor of five and the stress RMSE by roughly a factor of three.

\subsubsection{CrN benchmark}

\begin{figure*}[t]
    \centering
    \begin{subfigure}{0.49\textwidth}
        \centering
        \includegraphics[width=\textwidth]{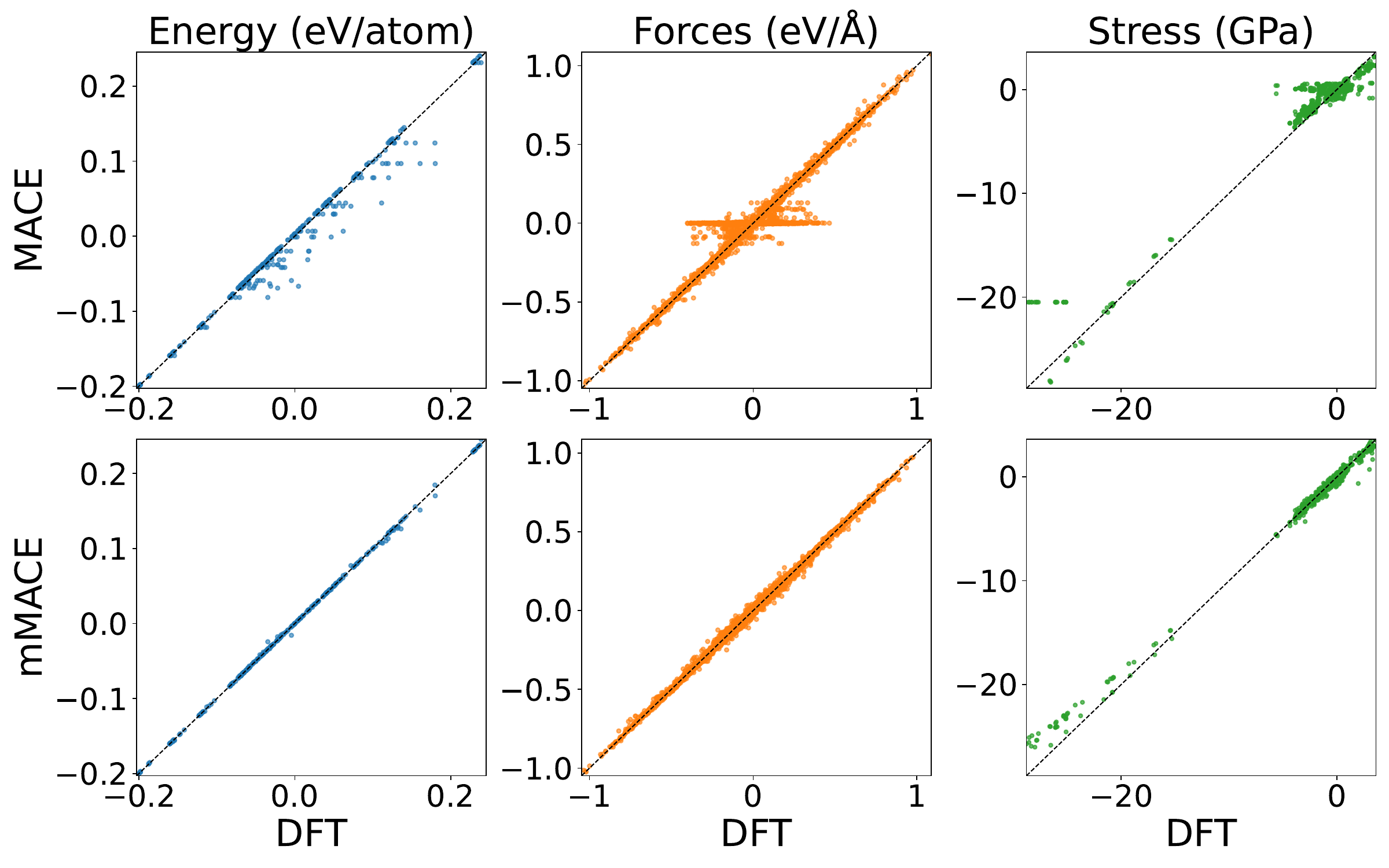}
        \caption{FeAl}
        \label{fig:scatter_feal}
    \end{subfigure}
    \hfill
    \begin{subfigure}{0.49\textwidth}
        \centering
        \includegraphics[width=\textwidth]{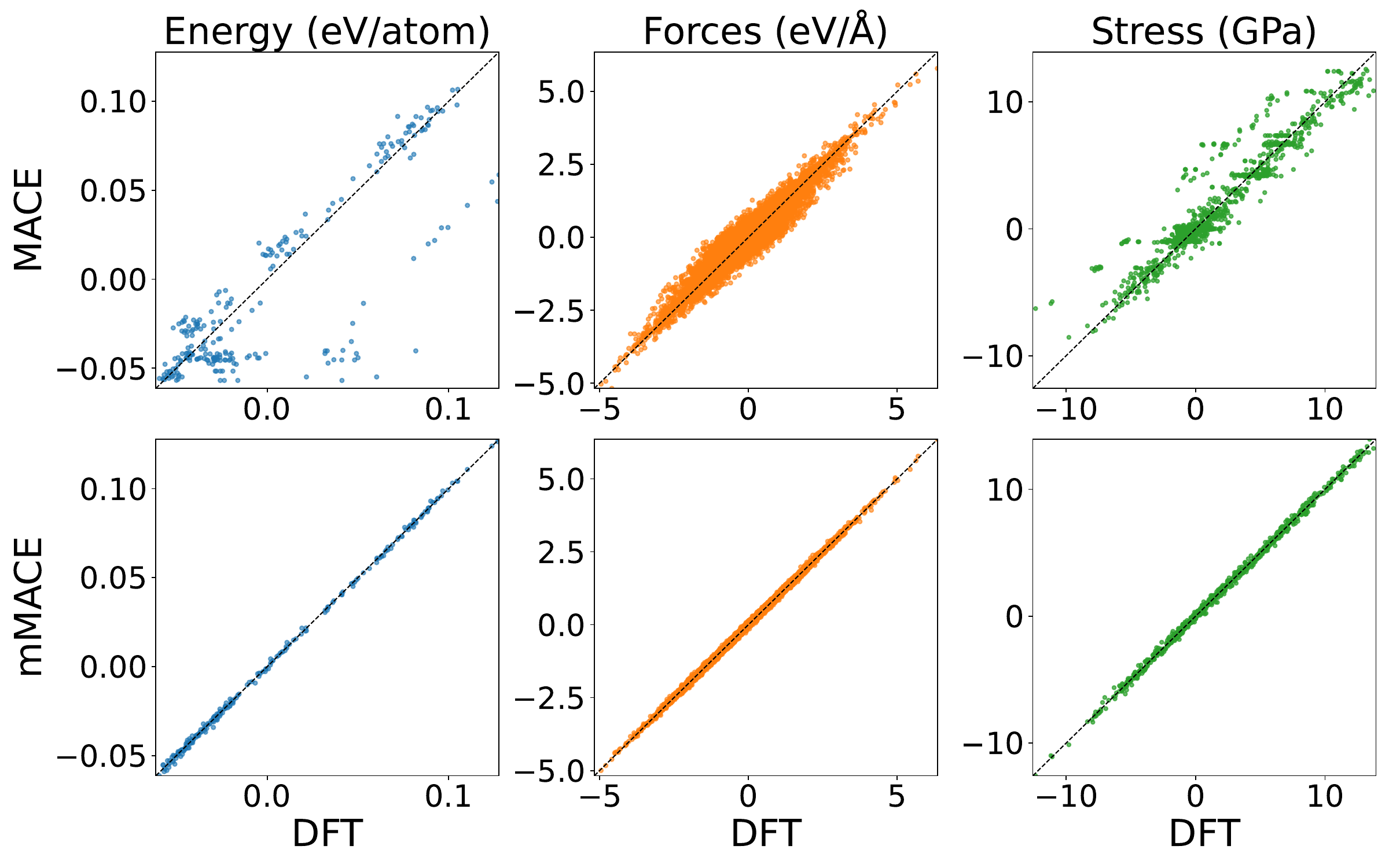}
        \caption{CrN}
        \label{fig:scatter_crn}
    \end{subfigure}

    \begin{subfigure}{0.48\textwidth}
        \centering
        \begin{tabular}{lccc}
        \hline
         & \textbf{mMTP} & \textbf{MACE} & \textbf{mMACE} \\
        \hline
        Energy (meV/atom) & 4.03 & 13.86 & \textbf{2.69} \\
        Force (meV/\AA)   & 48   & 52    & \textbf{11} \\
        Stress (GPa)      & 0.79 & 0.93  & \textbf{0.26} \\
        Magnetic force (meV/$\mu_B$) & -- & -- & -- \\
        \hline
        \end{tabular}
        \caption{FeAl RMSE test errors}
    \end{subfigure}
    \hfill
    \begin{subfigure}{0.48\textwidth}
        \centering
        \begin{tabular}{lccc}
        \hline
         & \textbf{mMTP} & \textbf{MACE} & \textbf{mMACE} \\
        \hline
        Energy (meV/atom) & 1.7  & 31.2 & \textbf{1.21} \\
        Forces (meV/\AA)  & 108  & 219  & \textbf{24} \\
        Stress (GPa)      & 0.4  & 1.29 & \textbf{0.25} \\
        Magnetic force (meV/$\mu_B$) & 64 & -- & \textbf{19} \\
        \hline
        \end{tabular}
        \caption{CrN RMSE test errors}
    \end{subfigure}

    \caption{
        Summary of benchmark results.  
        (a) Scatter plots comparing mMACE and MACE to DFT reference on the FeAl and CrN datasets.  
        (b–c) Corresponding RMSE values on the test sets \cite{kotykhov2025actively, kotykhov2023constrained}. A qualitative plot on magnetic forces of the CrN dataset can be found in Appendix~\ref{app:sec:CrN_mMACE_magforce_hist2d}. In contrast, the standard MACE model cannot predict magnetic forces at all, as this functionality is not included.
    }
    \label{fig:benchmark_with_tables}
\end{figure*}

Figure~\ref{fig:benchmark_with_tables}b presents the analogous comparison for the collinear spin-polarized CrN dataset of Kotykhov \textit{et al.}~\cite{kotykhov2025actively}. Here, the standard MACE model fails to resolve energy differences associated with different spin states, whereas mMACE successfully captures these distinctions and shows improved correlation with the DFT reference across energies, forces, and stresses.

The error metrics in Fig.~\ref{fig:benchmark_with_tables}b,d confirm that mMACE achieves the lowest errors across all quantities. Most notably, this dataset includes magnetic force labels, and the magnetic force RMSE is reduced by more than a factor of three relative to the mMTP baseline. A scatter plot of the magnetic force predictions is provided in Appendix~\ref{app:sec:CrN_mMACE_magforce_hist2d}.

\subsection{Pretrained Collinear Magnetic Models}
\label{sec:pretrained_model_training}

To demonstrate the ability of mMACE to fit broad, general-purpose datasets, we train on two publicly available sources that include magnetic information from collinear spin-polarized DFT: MATPES~\cite{kaplan2025foundational}, which provides both PBE and r\textsuperscript{2}SCAN reference calculations~\cite{perdew1996generalized}, and MP-ALOE~\cite{kuner2025mp}, which contains r\textsuperscript{2}SCAN  data. We fit mMACE models to each dataset independently and use the MATPES-PBE model as the starting point for the fine-tuning studies in Sections~\ref{sec:FeNi binary system} and~\ref{sec:Mn3Pt}.

The results are summarized in Table~\ref{table:pre_trained_errs}. On the MATPES-PBE dataset, mMACE consistently achieves lower mean absolute errors across energies, forces, and stresses compared to the baseline MACE model for configurations with non-zero magnetic moments. For zero-moment systems, where spin effects are minimal, mMACE maintains comparable performance, confirming that the additional magnetic degrees of freedom improve accuracy on magnetically active configurations without degrading predictions elsewhere.

\begin{table*}[t]
\label{tab:mae_all}
\renewcommand{\arraystretch}{1.2}
\setlength{\tabcolsep}{8pt}
\begin{tabular}{l l ccc ccc}
\hline
\multirow{2}{*}{\textbf{Dataset}} & 
\multirow{2}{*}{\textbf{Model}} &
\multicolumn{3}{c}{\textbf{Non-zero magmom}} &
\multicolumn{3}{c}{\textbf{Zero magmom}} \\
\cline{3-8}
 & & $E$ & $F$ & $S$ & $E$ & $F$ & $S$ \\
\hline
\multirow{2}{*}{MATPES–PBE} 
 & MACE  & 44 & 123 & 0.908 & \textbf{\underline{35}} & \textbf{\underline{92}} & \textbf{\underline{0.897}} \\
 & mMACE & \textbf{\underline{34}} & \textbf{\underline{119}} & \textbf{\underline{0.808}} & 39 & 96 & 0.947 \\
\hline
\multirow{2}{*}{MATPES–r2SCAN} 
 & MACE  & 47 & 154 & 1.040 & 41 & 117 & 1.117 \\
 & mMACE & \textbf{\underline{35}} & \textbf{\underline{133}} & \textbf{\underline{0.853}} & \textbf{\underline{36}} & \textbf{\underline{106}} & \textbf{\underline{1.003}} \\
\hline
\multirow{2}{*}{MP–ALOE} 
 & MACE  & 57 & 96 & 1.362 & 53 & 92 & 1.430 \\
 & mMACE & \textbf{\underline{48}} & \textbf{\underline{86}} & \textbf{\underline{1.133}} & \textbf{\underline{50}} & \textbf{\underline{88}} & \textbf{\underline{1.358}} \\
\hline
\end{tabular}
\centering
\caption{Mean absolute errors on the test set of mMACE and MACE across three datasets. 
$E$, $F$, and $S$ correspond to energy (meV/atom), forces (meV/\AA), and stress (GPa).}
\label{table:pre_trained_errs}
\end{table*}

The improvement is consistent across all three datasets, though more modest than the five- to tenfold error reductions observed on the FeAl and CrN benchmarks (Section~\ref{sec:public_datasets}). This is expected: MATPES and MP-ALOE are dominated by ferromagnetic configurations, so the magnetic diversity against which mMACE can differentiate itself is limited. Nevertheless, the pre-trained mMACE model serves as a reliable baseline for fine-tuning on specific magnetic systems, as we demonstrate in Sections~\ref{sec:FeNi binary system} and~\ref{sec:Mn3Pt}.

We note that single-point error metrics do not fully reflect a model’s ability to reproduce self-consistent magnetic states under energy minimization. An analysis of relaxed magnetic moments and its implications for the stability and transferability of magnetic MLIPs is provided in Appendix~\ref{app:relaxed_magnetic_moments}.

\subsection{Fine-tuning study: \ce{FeNi} Binary System}
\label{sec:FeNi binary system}

We next apply mMACE to the \ce{FeNi} binary alloy, a prototypical transition-metal system of structural and magnetic relevance \cite{edstrom2014electronic}. This section serves two purposes: (i) to assess whether mMACE can resolve magnetically sensitive phenomena, in particular the Bain transformation connecting the fcc and bcc phases; and (ii) to demonstrate a data-efficient fine-tuning strategy starting from the MATPES-PBE pre-trained model.

\begin{figure*}[t]
    \centering
    
    \begin{subfigure}[t]{0.48\textwidth}
        \centering
        \vspace{-8.5cm}
        \begin{subfigure}[t]{\linewidth}
            \centering
            \includegraphics[width=\linewidth]{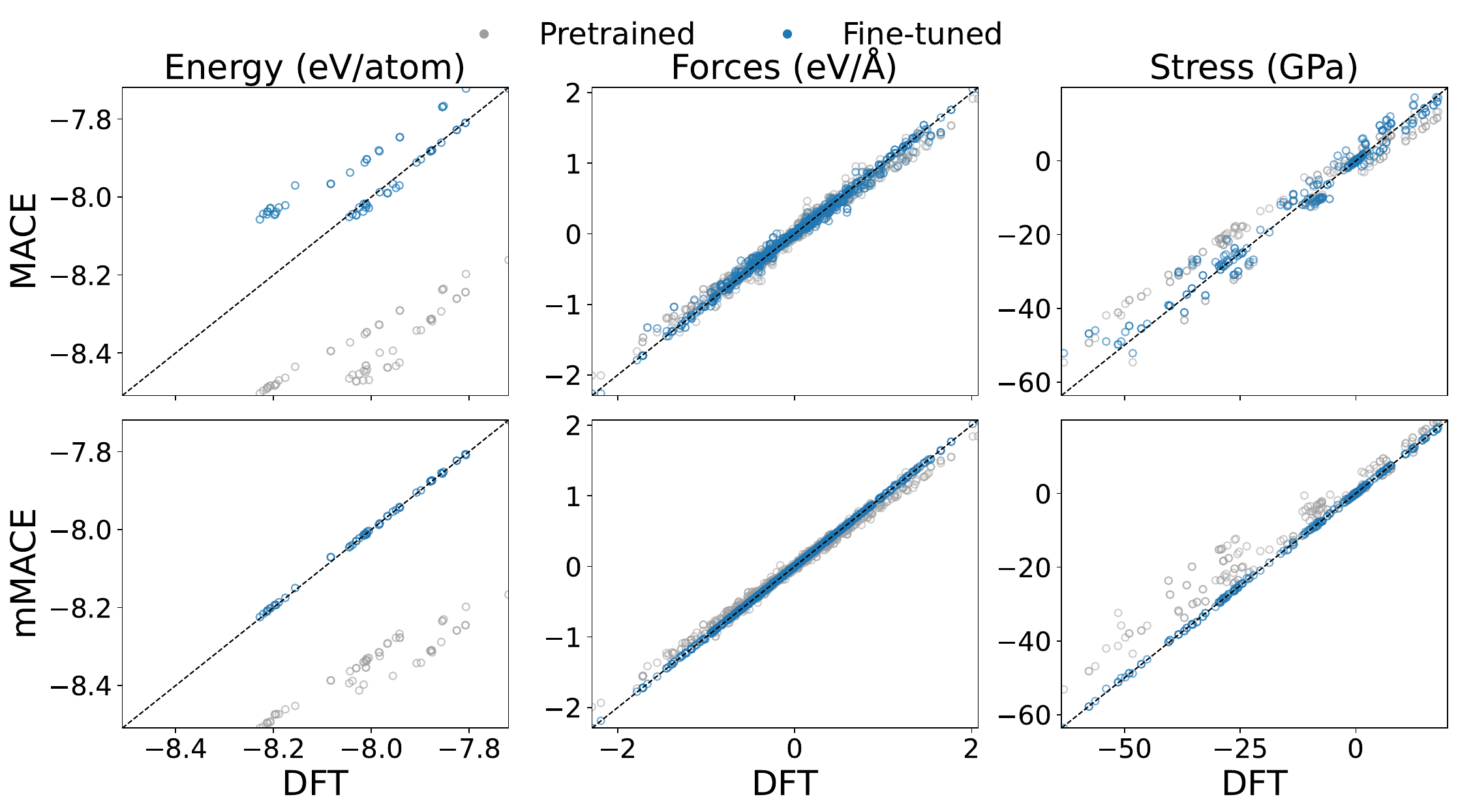}
            \caption{}
            \label{fig:FeNi_error}
        \end{subfigure}
        
        \vspace{6pt}
        
        \begin{subfigure}[t]{\linewidth}
            \centering
            \begin{tabular}{lccc}
            \hline
            \textbf{Model} 
            & Energy 
            & Forces 
            & Stress \\
            & (meV/atom) 
            & (meV/\AA) 
            & (GPa) \\
            \hline
            MACE-MATPES-PBE     
            & 382.7 
            & 69.76 
            & 2.479 \\
            
            mMACE-MATPES-PBE      
            & 349.8 
            & 54.57 
            & 2.552 \\
            
            MACE-FT (iter3)   
            & 66.7 
            & 37.14 
            & 1.392 \\
            
            mMACE-FT (iter3)       
            & \textbf{\underline{2.44}} 
            & \textbf{\underline{3.06}} 
            & \textbf{\underline{0.117}} \\
            \hline
            \end{tabular}
            \caption{}
            \label{tab:finetune_comparison}
        \end{subfigure}
    \end{subfigure}
    \hfill
    \begin{subfigure}[t]{0.48\textwidth}
        \centering
        \includegraphics[width=\linewidth]{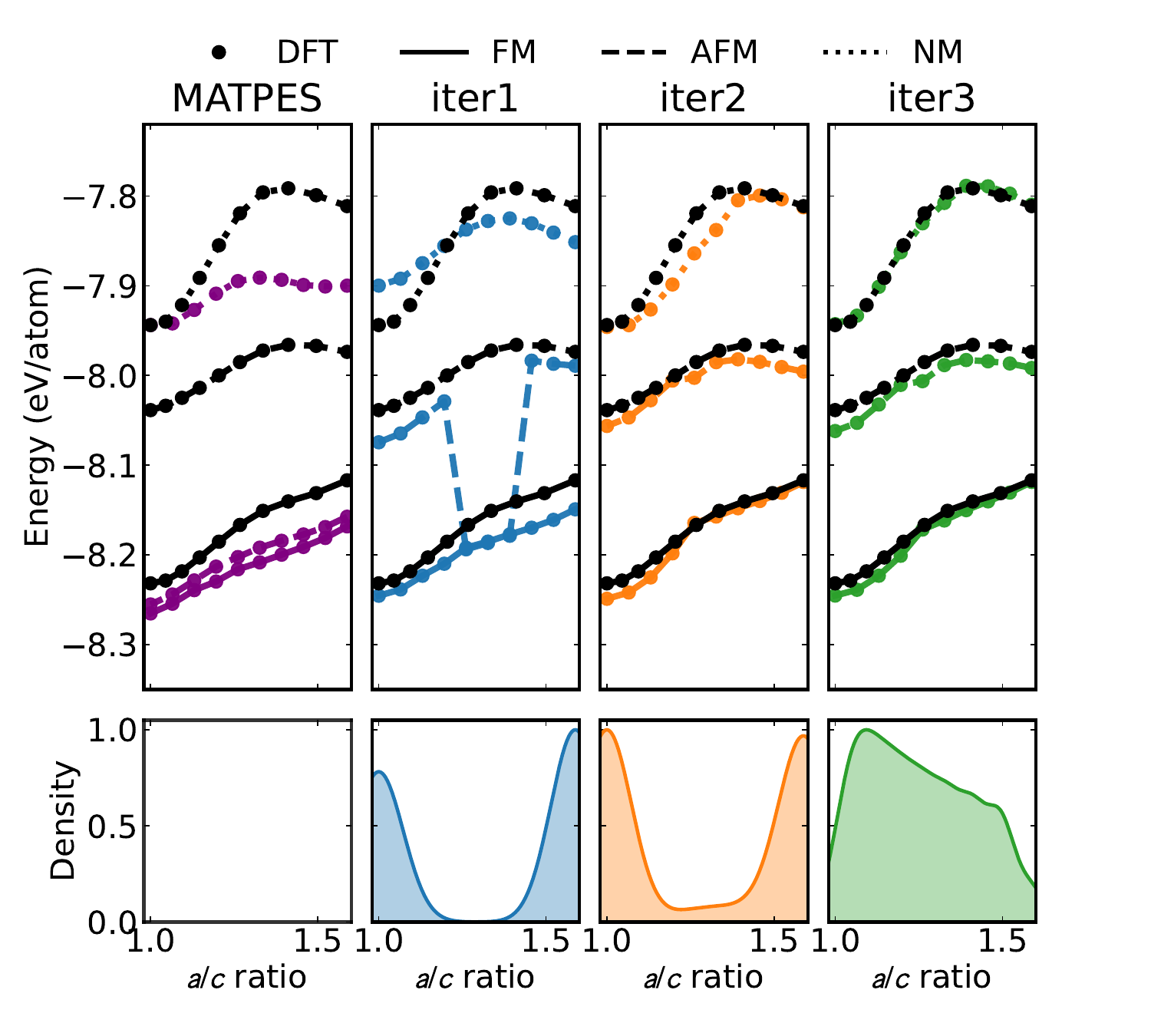}
        \caption{}
        \label{fig:FeNi_BP_all}
    \end{subfigure}
    
    \caption{
    Finetuning of the FeNi database using mMACE. 
    \textbf{(a, b) Left:} Energy, force and stress predictions of (m)MACE models against DFT reference with MAE values on the FeNi finetuning database. Errors after energy minimization (with DFT magnetic moment as initial guess) with mMACE-FT (iter3) are 34.6 meV/atom (energy), 18.2 meV/\AA\ (forces), and 0.883 GPa (stress). 
    \textbf{(c) Right:} mMACE Bain path transformation for FM, AFM, and NM states as the dataset is iteratively enriched (Section~\ref{sec:FeNi binary system}). Density plots indicates the distribution of additional data used in the face-centered tetragonal representation in each iteration, where $a/c = 1$ and $a/c = \sqrt{2} \approx 1.414$ corresponds to fcc and bcc respectively. Energies of mMACE-MATPES are shifted by matching the $a/c=1.0$ NM state to remove the constant offset arising from differences between DFT implementations for comparison between different magnetic states.
    }
    \label{fig:FeNi_combined}
\end{figure*}

Before constructing a dedicated dataset, we assess the pre-trained model on the Bain path. Figure~\ref{fig:FeNi_BP_all} shows that it correctly captures the energy scale separating non-magnetic and ferromagnetic states along the bcc--fcc transformation. However, it fails to resolve the relative ordering of the ferromagnetic and antiferromagnetic configurations. This is expected, as only 1,380 structures (0.35\%) of the MATPES training dataset are antiferromagnetic, as classified by the \texttt{CollinearMagneticStructureAnalyzer} implemented in \texttt{pymatgen}~\cite{Ong2013pymatgen}, version 2025.10.7.

To address this, we generate targeted \ce{FeNi} configurations using the protocol in Appendix~\ref{app:sec:data_generation:FeNi}, progressively enriching structural coverage along the corresponding cell ratio of interest without explicitly sampling the Bain path (see the lower panels Fig.~\ref{fig:FeNi_BP_all}). This results in a sequence of refined models with systematically improved accuracy. 

The resulting Bain path after the second fine-tuning iteration (iter2 in Fig.~\ref{fig:FeNi_BP_all} 
already matches the DFT reference almost exactly, even though no explicit Bain-path configurations were included in the training set. This provides strong evidence that mMACE is sufficiently expressive to capture the physics governing this transformation from generic structural data alone.

To further assess the mMACE model, we perform an additional fine-tuning stage (labeled as iter3) in which a small number of explicitly computed Bain-path configurations is incorporated into the training set. Following this fine-tuning step, the accuracy along all three transformation paths improves, with the most pronounced gains observed for the NM transformation. This demonstrates that mMACE can fit a broad range of configurations, provided the training data adequately span the relevant structural and magnetic degrees of freedom. As summarized in Table~\ref{tab:finetune_comparison}, the fine-tuned mMACE model significantly outperforms the baseline MACE-MATPES-PBE model across energy, force, and stress metrics. This improvement highlights the ability of mMACE to capture the underlying magnetic and structural physics with substantially higher fidelity.

Without introducing further training data, we additionally evaluate elastic constants. All equilibrium lattice parameters agree with DFT to within 0.01~\AA\ across all phases and magnetic states, and all elastic constants reproduce the correct trends. Further details are provided in Appendix~\ref{apd:sec:feni_lattice_parameters}.

Overall, these results demonstrate that magnetically sensitive phase stability and mechanical properties in \ce{FeNi} can be recovered through targeted, physically motivated fine-tuning of a pre-trained magnetic model, without large-scale retraining.

\subsection{Fine-tuning study: non-collinear \ce{Mn3Pt} }
\label{sec:Mn3Pt}

To evaluate mMACE on non-collinear magnetism, we consider the frustrated antiferromagnet \ce{Mn3Pt}, which features a triangular spin texture on a kagome-like Mn sublattice~\cite{zuniga2023observation}. Strong geometric frustration and competing exchange interactions stabilize a non-collinear ground state, making \ce{Mn3Pt} a demanding test case: resolving the small energy differences between competing magnetic configurations requires an accurate treatment of spin-dependent interactions beyond collinearity.

The data generation procedure detailed in Appendix~\ref{app:sec:Mn3Pt_datagen} yields 147 training and 100 validation configurations. We use this dataset to fine-tune the MATPES-PBE pre-trained mMACE model, with the goal of robustly recovering the kagome-like ground state as the energy minimum.

\begin{figure*}[t]
   \centering
    \begin{subfigure}[t]{0.28\textwidth}
        \centering
        \includegraphics[height=4.2cm]{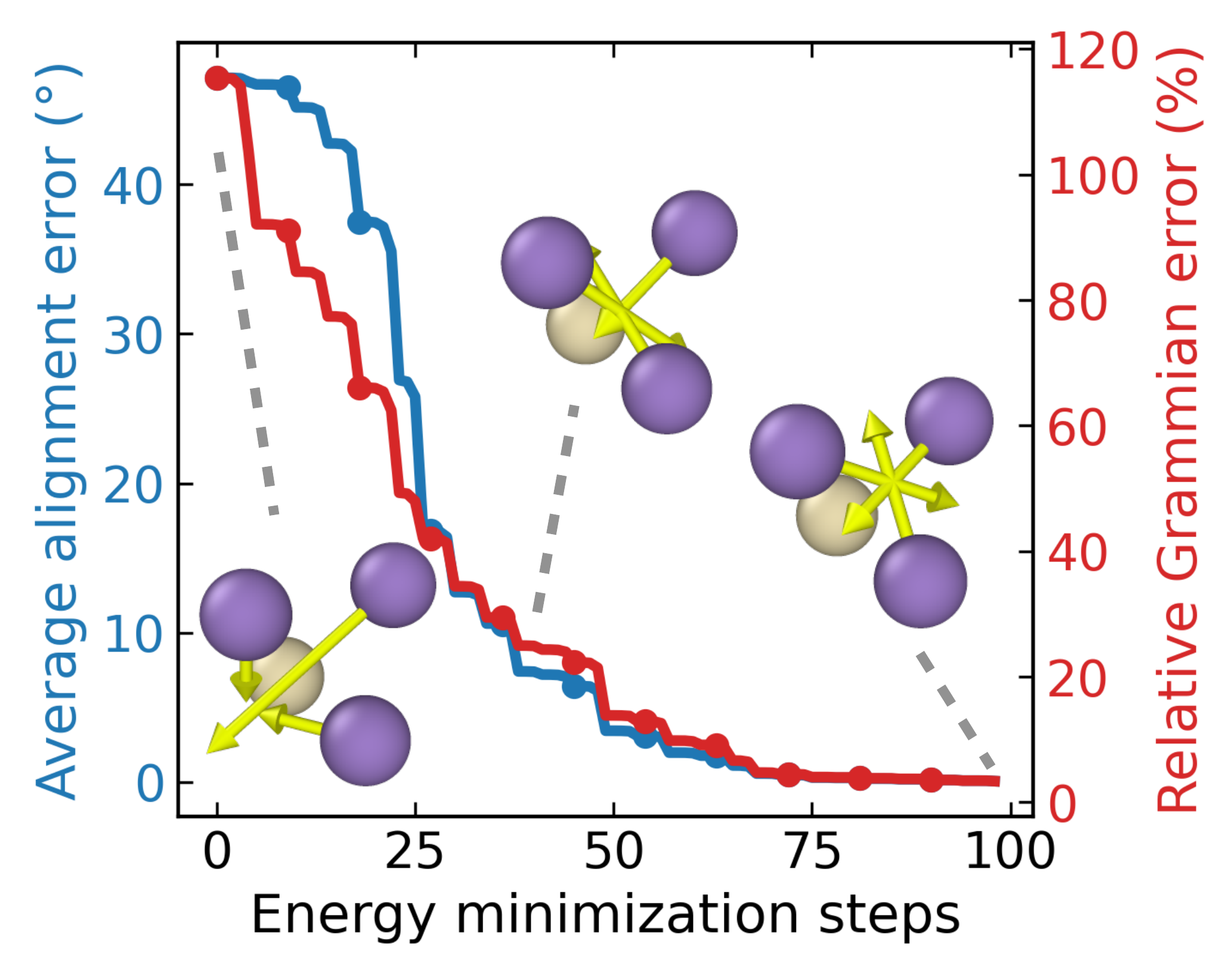}
        \caption{}
        \label{fig:Mn3Pt_energy_min}
    \end{subfigure}
    \hfill
    \begin{subfigure}[t]{0.31\textwidth}
        \centering
        \includegraphics[height=4.2cm]{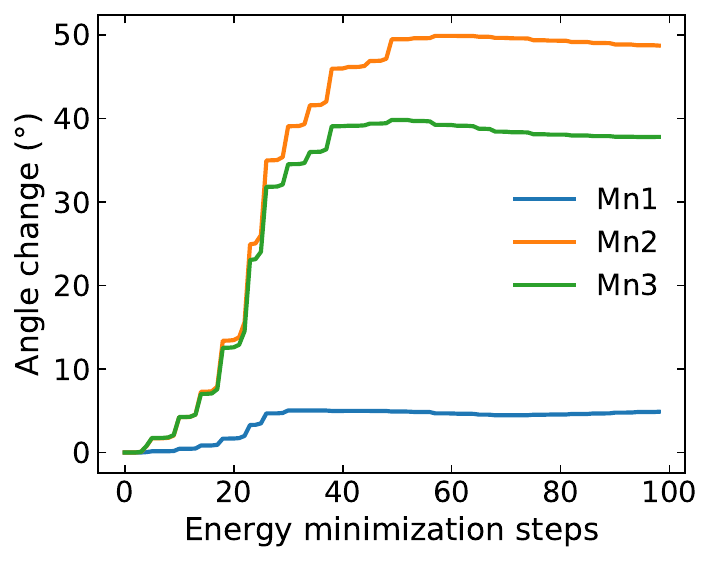}
        \caption{}
        \label{fig:Mn3Pt_energy_min_angle}
    \end{subfigure}
    \hfill
    \begin{subfigure}[t]{0.31\textwidth}
        \centering
        \includegraphics[height=4.2cm]{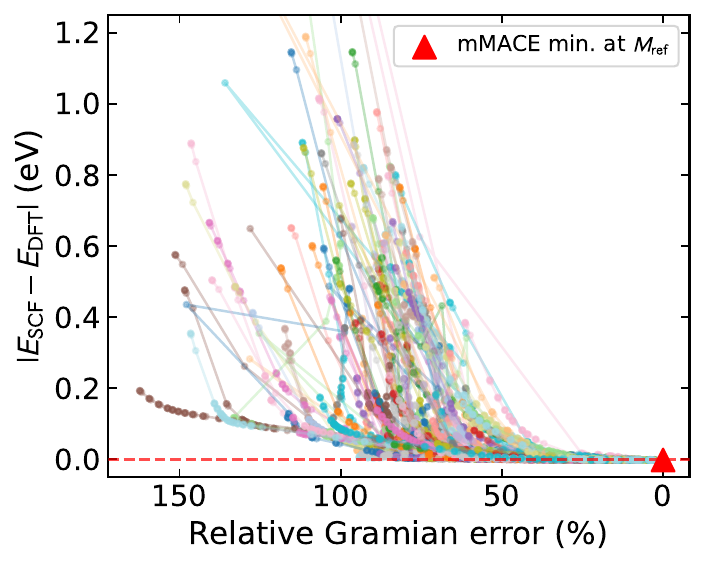}
        \caption{}
        \label{fig:Mn3Pt_scf_misalignment}
    \end{subfigure}

    \caption{(a) Energy minimization of magnetic moments on the Mn$_3$Pt lattice starting from a random orientation, ultimately relaxing to the kagome-like spin structure. (b) Rotation angle relative to a reference orientation as a function of minimization steps. Each energy minimization step corresponds to a single energy evaluation.
    (c) Energy error with respect to ground state decreases monotonically as the magnetic moments become better aligned. Large misalignments produce substantial energy residue. The red triangle indicates the mMACE-relaxed magnetic configuration, which lies at zero misalignment and zero energy residue.}
    \label{fig:Mn3Pt_all}
\end{figure*}

In the absence of SOC, magnetic configurations are symmetry equivalent under global spin rotations. To define an error metric that is invariant under this symmetry, we consider the relative Gramian error for a matrix $A$ containing stacked magnetic moment vectors 
\begin{equation}
\varepsilon_{\mathrm{rel}}(A, A_{\rm ref})
=
\frac{\left\| G(A) - G(A_{\rm ref}) \right\|_F}
{\left\| G(A_{\rm ref}) \right\|_F}.
\end{equation}
where $G(A) := A^T A$ is the Gramian matrix and $\| \cdot \|_{F}$ is the Frobenius norm. As shown in Fig.~\ref{fig:Mn3Pt_energy_min}, both the Gramian error and the angle alignment error decrease rapidly during the energy minimization and converge to near-zero values, confirming that the optimized spin configuration matches the kagome-like ground
state up to a global rotation. Figure~\ref{fig:Mn3Pt_energy_min_angle} illustrates the corresponding evolution of the Mn magnetic moment orientations, which undergo substantial rotations from their random initial values before stabilizing into the characteristic frustrated spin arrangement.

To demonstrate that mMACE can navigate the highly non-convex magnetic energy landscape, we perform 100 relaxations from random initial spin orientations, all outside the training and validation distributions. Figure~\ref{fig:Mn3Pt_scf_misalignment} shows that the energy residue decreases systematically with decreasing magnetic misalignment, and that all runs converge to the kagome-like ground state. The MACE reference point (red triangle), relaxed from the DFT ground state, lies at zero misalignment and negligible energy residual, confirming that magnetic relaxation restores a well-converged self-consistent solution. Details on the relaxation procedure are provided in Appendix~\ref{app:sec:Mn3Pt_relaxations}.

\subsection{Case study: magnetocrystalline anisotropy}
\label{sec:magnetocrystalline_anisotropy}

Magnetocrystalline anisotropy (MCA) describes the directional preference of spin alignment along specific crystallographic axes. It originates from the coupling between the electronic spin and the crystal lattice \textit{via} spin--orbit coupling (SOC)~\cite{miura2022understanding, le2019magnetocrystalline}.

For a fixed atomic configuration, the magnetic anisotropy energy $E_{A}$ is defined as the total energy relative to a reference magnetization orientation $(\theta_0, \phi_0)$,
\begin{equation}
    E_{A}(\theta, \phi) 
    := E(\theta, \phi) - E(\theta_0, \phi_0),
\end{equation}
where $(\theta, \phi)$ denote the polar and azimuthal angles of the magnetization vector. The orientation that minimizes the energy defines the \textit{easy axis} of the system.

Typically, the energy scale associated with SOC-induced anisotropy is on the order of sub-meV per atom (see Fig.~\ref{fig:soc_distribution}), which is comparable to or smaller than the intrinsic error of modern machine-learned interatomic potentials in single-point energy predictions. Consequently, the anisotropy should be regarded as a small perturbative correction to the total energy of a configuration and is fitted separately. 

In this work, we are primarily interested in identifying configurations that exhibit a strong anisotropic response to magnetization reorientation. To quantify this, we define the anisotropy strength of a configuration $\{\sigma_i\}_i := \{(\mathbf{r}_i, \mathbf{m}_i, z_i)\}_i$ as the maximal variation of the total energy over all magnetization directions,
\begin{align}
\Delta E_{A}(\{\sigma_i\}_i)
&:= \max_{\theta_1,\phi_1,\theta_2,\phi_2}
\Big[
E(\theta_1,\phi_1;\{\sigma_i\}_i)
\nonumber\\
&\hspace{3.5em}
-
E(\theta_2,\phi_2;\{\sigma_i\}_i)
\Big].
\end{align}
This quantity characterizes the intrinsic anisotropy scale associated with a given atomic and magnetic configuration. Details of the dataset construction and computational protocol are provided in Appendix~\ref{app:sec:data_generation_magnetic_anitostropy}.

\subsubsection{\ce{FeMn}}
We consider a ferrimagnetic configuration of $L1_0$ \ce{FeMn} 
within a two-atom unit cell. SOC is included non-self-consistently in all 
calculations. The magnetic anisotropy energy  is obtained by rotating the spin axis while keeping the atomic positions fixed. For this uniaxial system, the magnetic anisotropy energy $E_{A}$ can be well approximated by a fourth-order angular expansion~\cite{collings2023generalized}
\begin{equation}
\frac{E_{A}}{V} \approx K_1 \sin^2\theta + K_2 \sin^4\theta.
\end{equation}
where $V$ denotes the system volume and the $z$-axis coincides with the principal symmetry axis. The sign of $K_1$ determines whether this axis corresponds to an \textit{easy} or \textit{hard} magnetization direction, and typically $|K_1| \gg |K_2|$. Details on dataset construction can be found in Appendix~\ref{app:sec:FeMn_soc_datagen}.

Figure~\ref{fig:uniaxial_fit} compares the angular dependence obtained 
from DFT and the mMACE model. The fitted values of $K_1$ are 
$2.20~{\rm Jm}^{-3}$ (mMACE) and $2.22~{\rm Jm}^{-3}$ (DFT), 
demonstrating excellent agreement. The fitted $K_2$ term is several 
orders of magnitude smaller than $K_1$ and cannot be reliably 
determined within numerical noise.

\begin{figure}[t]
    \centering
    \hspace{-12pt}
    \includegraphics[width=0.95\linewidth]{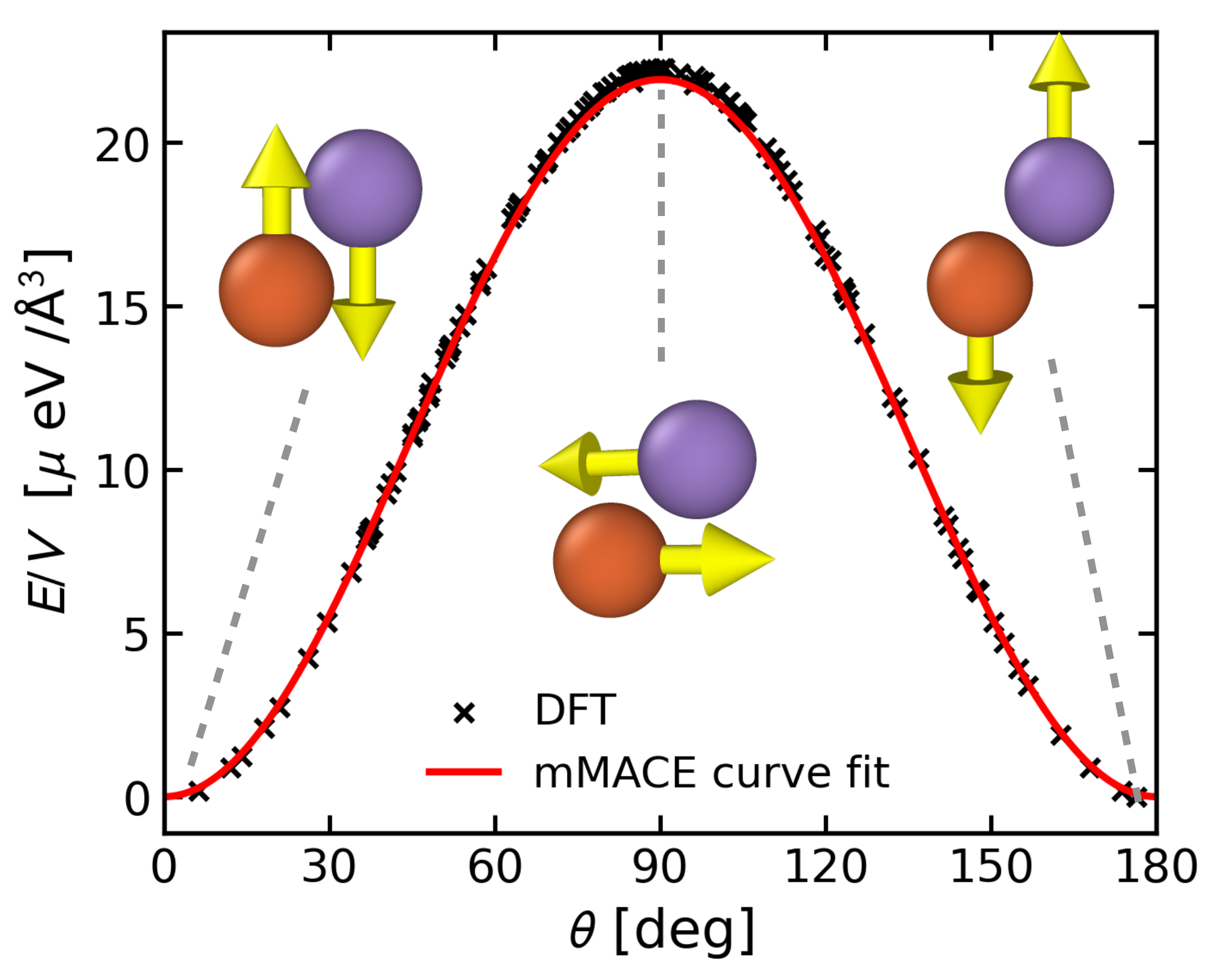}
    \vspace{-7pt}
    \caption{Comparison between DFT data and mMACE energy density curve for FeMn.}
    \label{fig:uniaxial_fit}
    \vspace{-8pt}
\end{figure}

\subsubsection{Fitting a random structure search (RSS) database}
%
%
To assess the generalization capability of mMACE in predicting $E_{A}$ across diverse chemical environments, we construct a database of random structures spanning a broad range of metallic compositions. The dataset includes transition and post-transition metals with varying strengths of spin--orbit coupling---namely Al, Mn, Fe, Co, Ni, Pt, and Y---and additionally incorporates up to one light element selected from B, C, or Si. A total of 242 \emph{parent structures} were created, and for each parent structure 2000 orientations were sampled over the unit sphere, resulting in 484k configurations.

\begin{figure*}[t]
    \centering

    \begin{subfigure}[t]{0.51\textwidth}
        \centering
        \includegraphics[width=\linewidth]{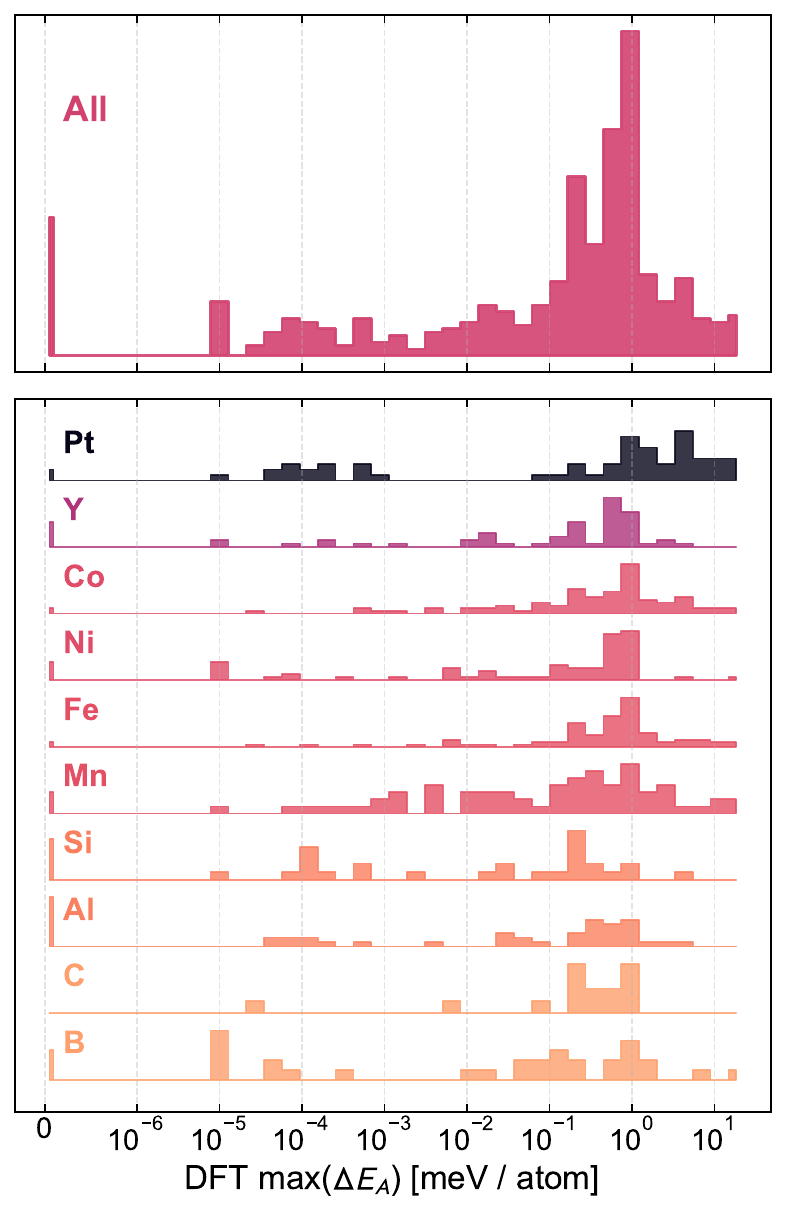}
        \caption{}
        \label{fig:soc_distribution}
    \end{subfigure}
    \hfill
    \begin{subfigure}[t]{0.39\textwidth}
        \centering
        \vspace{-14cm}
        \includegraphics[width=\linewidth]{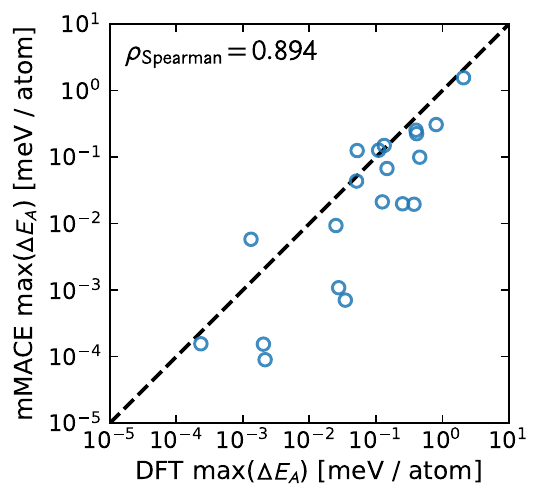}
        \caption{}
        \label{fig:correlation_max_soc_deltaE}
        \vspace{3pt}
        
        \includegraphics[width=\linewidth]{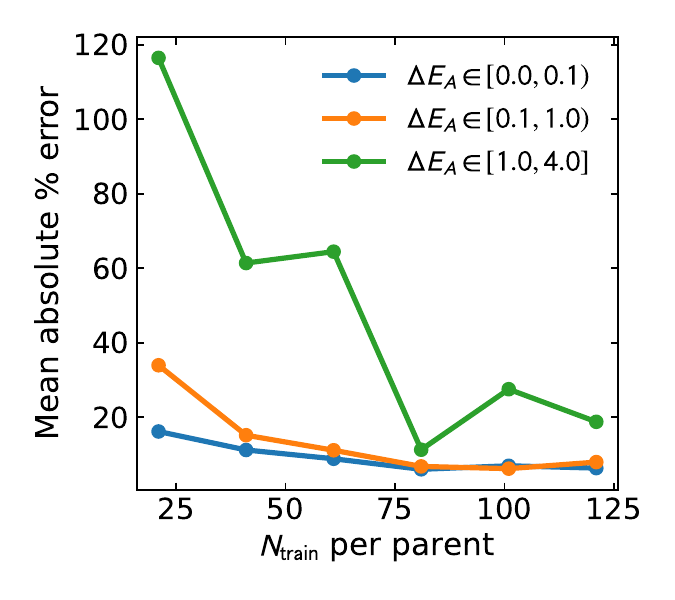}
        \caption{}
        \label{fig:soc_diff_learning_curve}
        \label{fig:soc_combined_right}
    \end{subfigure}

    \caption{
    Spin–orbit coupling (SOC) energy sensitivity across the RSS dataset.
    \textbf{Left:} Distribution of maximum SOC-induced magnetic anisotropy energies $\Delta E_{A}$ over all configurations containing a given element. Elements are ordered by median $\Delta E_{A}$, illustrating the broad dynamic range of SOC sensitivity.
    \textbf{Right:} (Top) Maximum SOC energy variation per atom predicted by mMACE versus DFT on logarithmic scales on a test set; the dashed line denotes perfect agreement. (Bottom) Learning curve for SOC energy differences with percentage error regularized by 1 $\mu$ eV / atom, showing systematic error reduction with increasing training data.
    }
    \label{fig:soc_all}
\end{figure*}

To account for the wide dynamic range of spin--orbit coupling energies, the results are subdivided by $E_{\rm SOC}$ range, following the distribution in Fig.~\ref{fig:soc_distribution}. Since $\Delta E_{A}$ spans several orders of magnitude, we report relative rather than absolute errors. We extract varying numbers of training configurations $N_{\rm train}$ per parent structure, using 190 configurations each for training and validation, with the remainder reserved for testing. The resulting learning curves are shown in Fig.~\ref{fig:soc_diff_learning_curve}, reporting the relative mean absolute error. The model converges to a relative error of approximately 5\%, although the relative error is larger for configurations with stronger spin--orbit coupling.

To assess the model’s ability to identify configurations with strong SOC, we perform a 90--10 train-test split at the level of \emph{parent configurations}, \textit{i.e.}\ all SOC-rotated structures derived from the same parent are assigned exclusively to either the training or the test set.

Figure~\ref{fig:correlation_max_soc_deltaE} compares the maximum SOC-induced energy $\Delta E_{A}$ per parent configuration on the test set, predicted by mMACE against the corresponding DFT reference values. Both axes are shown on logarithmic scales, and the dashed line indicates perfect agreement. The strong monotonic correlation between mMACE and DFT, quantified by a Spearman rank coefficient of $\rho = 0.894$, demonstrates that the model reliably identifies parent configurations associated with large SOC energy variations. While deviations in absolute magnitude remain, the relative ordering of SOC sensitivity across configurations is well preserved, which is the key requirement for SOC-aware configuration screening.

\subsection{Case study: non-collinear Fe model}
\label{sec:non_collinear_fe}

Rinaldi \textit{et al.}~\cite{rinaldi2024non} introduced a comprehensive dataset for pure Fe covering an extensive range of atomic configurations, including non-collinear magnetic data. The ACE model developed in that work was shown to reproduce a broad spectrum of physical properties with high accuracy, including Curie temperatures. We fit this Fe database with mMACE to benchmark its performance, and compare methods of Curie temperature estimation---in particular contrasting mMACE-driven simulations with a classical Heisenberg spin model---to highlight the benefits of magnetically informed ML potentials.

\vspace{-0.2cm}
\subsubsection{Heisenberg exchange constant}

A widely used framework for describing magnetism in condensed matter is the classical Heisenberg model
\begin{equation}
    \mathcal{H} := \mathcal{H}_{0} - \sum_{\langle ij \rangle} J_{ij} \hat{\mathbf{S}}_{i} \cdot \hat{\mathbf{S}}_{j}
\end{equation}
where $\hat{\mathbf{S}}_{i}$ are normalized spin vectors and $J_{ij}$ are exchange coefficients between pairs of atomic sites. It has been actively used in magnetic materials research and the development of electronic devices~\cite{mi2023two}.

\begin{figure*}[t]
    \centering

    \begin{subfigure}{0.48\textwidth}
        \centering
        \includegraphics[width=\linewidth]{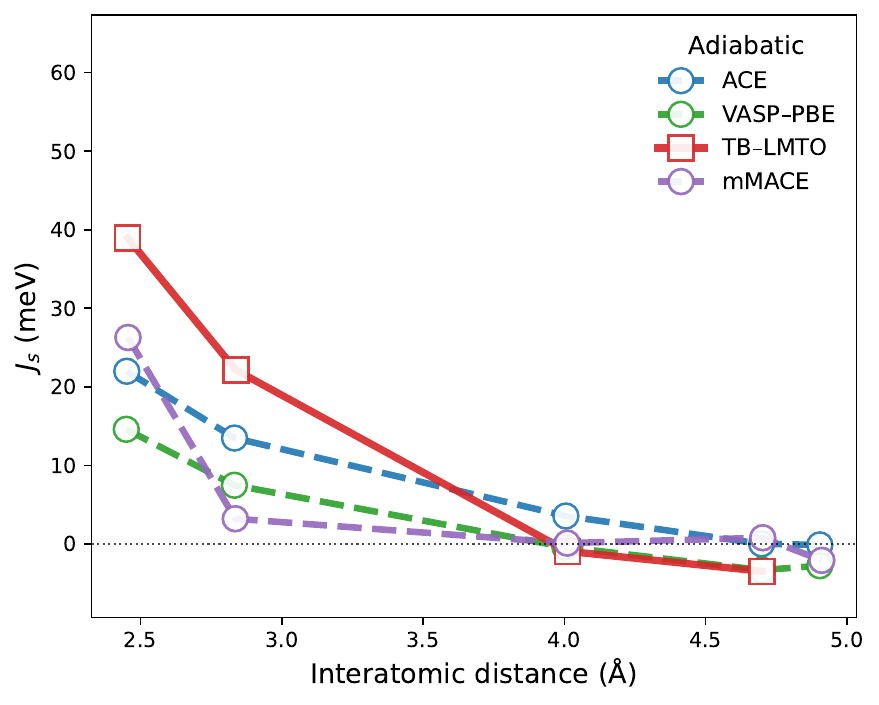}
        \caption{Heisenberg exchange constants obtained using the real-space approach of Liechtenstein et al.\ \cite{liechtenstein1984exchange, liechtenstein1986lsdf, liechtenstein1987local}.}
        \label{fig:J_adiabatic}
    \end{subfigure}
    \hfill
    \begin{subfigure}{0.48\textwidth}
        \centering
        \includegraphics[width=\linewidth]{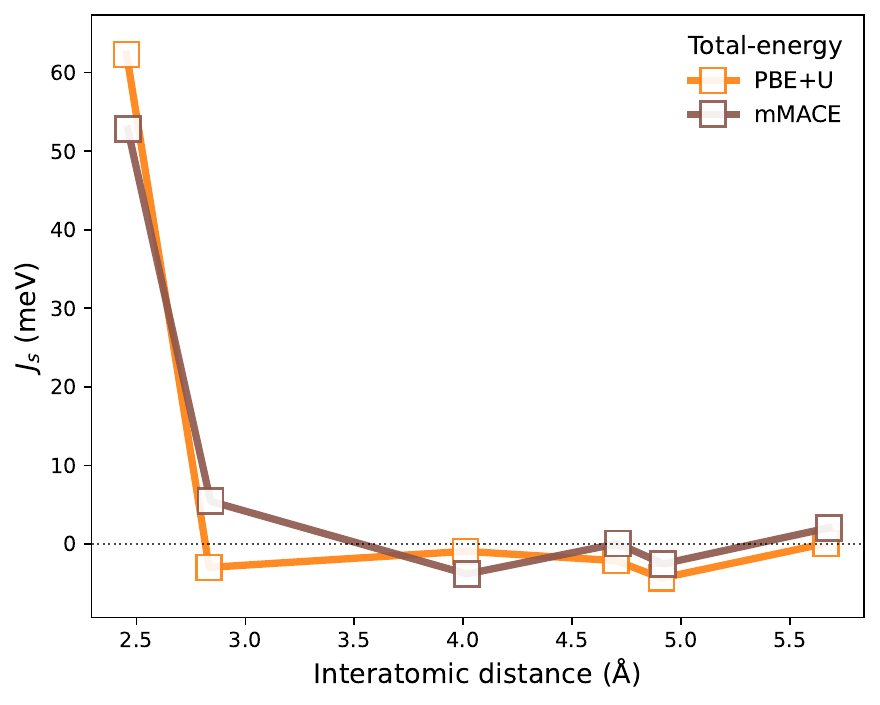}
        \caption{Heisenberg exchange constants obtained using the total-energy method in \cite{priessnitz2025ostravaj}.}
        \label{fig:J_total}
    \end{subfigure}

    \vspace{1em}

    \begin{subfigure}{0.48\textwidth}
        \centering
        \includegraphics[width=\linewidth]{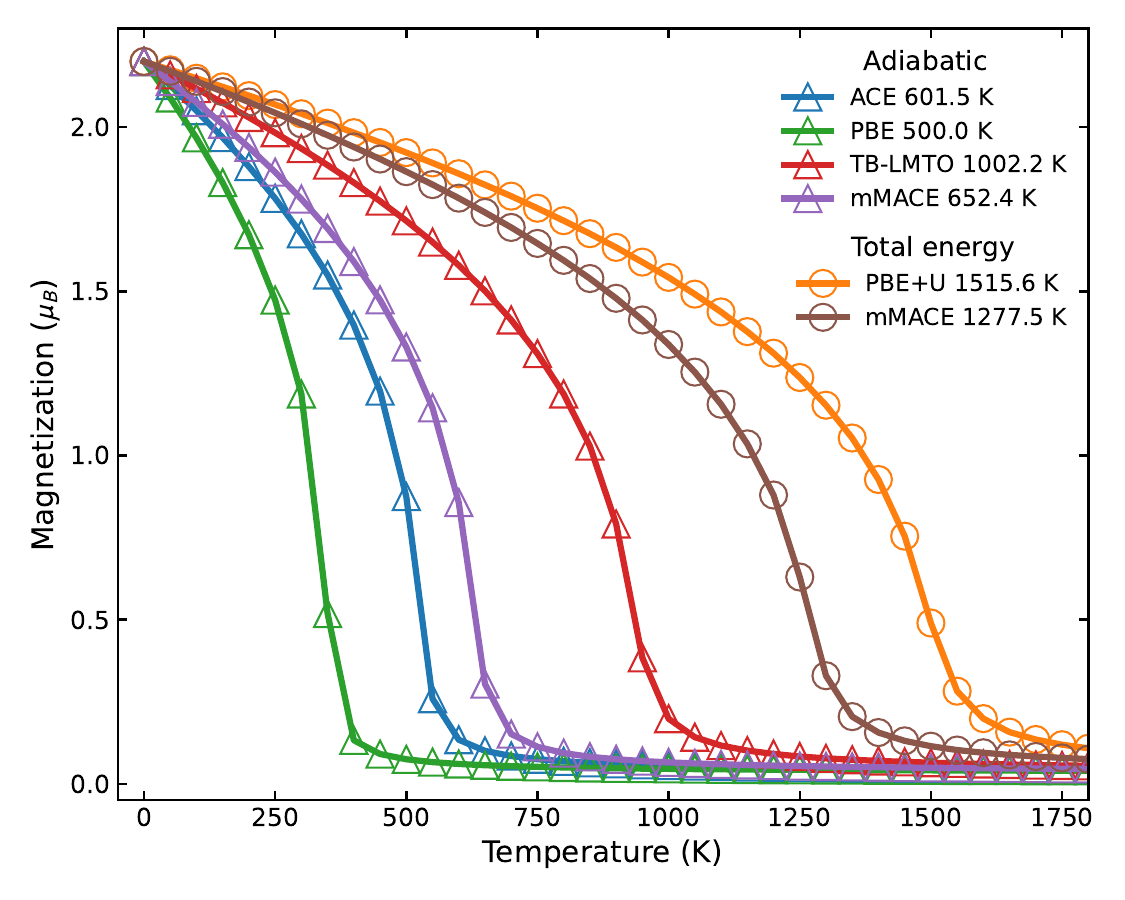}
        \caption{VAMPIRE MC simulations with $J_{ij}$ from the adiabatic and total-energy methods.}
        \label{fig:curie_vampire}
    \end{subfigure}
    \hfill
    \begin{subfigure}{0.48\textwidth}
        \centering
        \includegraphics[width=\linewidth]{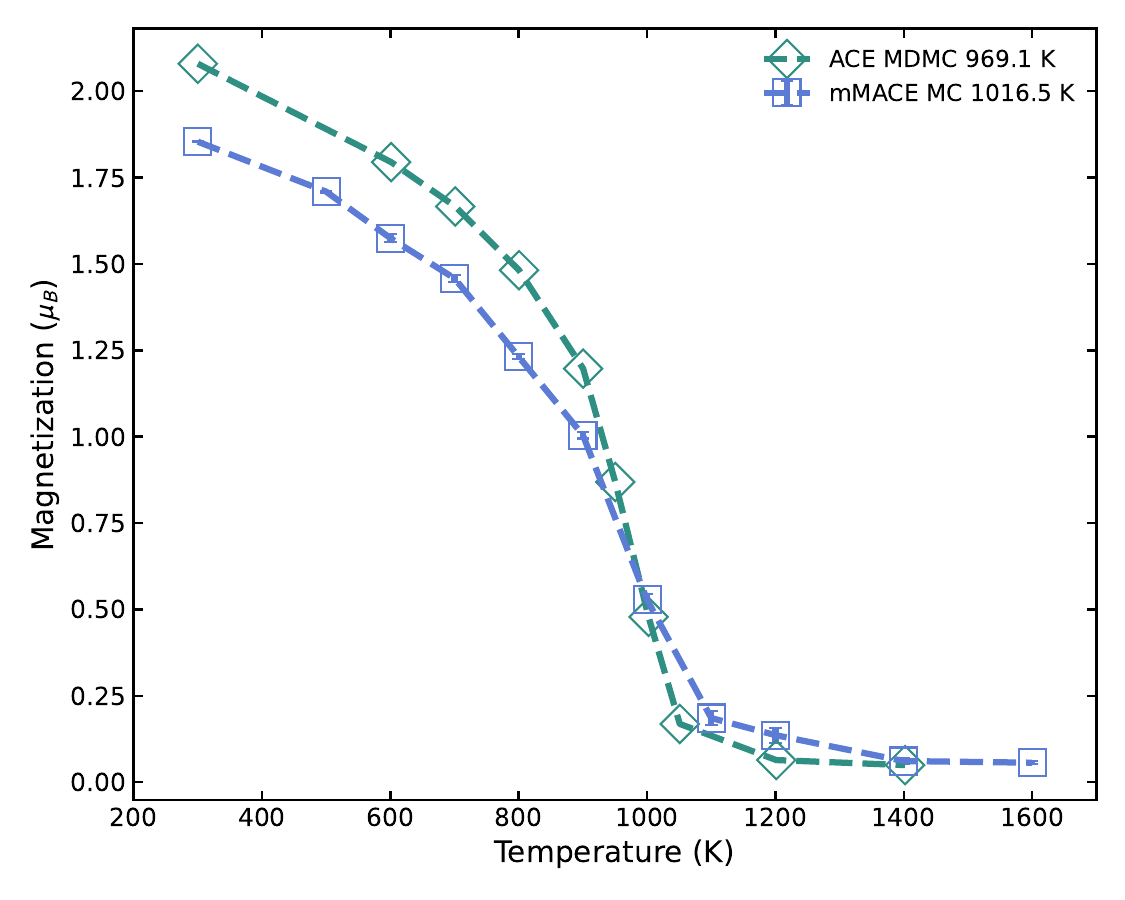}
        \caption{MD–MC and MC simulations using different ACE and mMACE. ACE reference is obtained from \cite{rinaldi2024non}.}
        \label{fig:curie_mdmc}
    \end{subfigure}

    \caption{
    Heisenberg exchange interactions and resulting magnetic phase transitions in bcc Fe.
    (a,b) Variation of Heisenberg exchange constants $J_s$ as a function of interatomic distance obtained using the adiabatic and total-energy methods, respectively. VASP–PBE reference data are taken from Ref.~\cite{rinaldi2023modelling}. (c,d) Temperature-dependent magnetisation curves showing the ferromagnetic–paramagnetic phase transition around the Curie temperature, computed using VAMPIRE Monte Carlo simulations and MD--MC simulations with machine-learned interatomic potentials.
    }
    \label{fig:heisenberg_curie_combined}
\end{figure*}

Two widely used approaches for estimating Heisenberg exchange coefficients are the total-energy--difference method~\cite{priessnitz2025ostravaj} (see Appendix~\ref{app:sec:heisenberg_exchange_constant} for details) and the real-space approach of Liechtenstein \textit{et al.}~\cite{liechtenstein1984exchange, liechtenstein1986lsdf, liechtenstein1987local}. These fundamentally different approaches yield qualitatively distinct estimates of the exchange constants. Figures~\ref{fig:J_adiabatic} and~\ref{fig:J_total} present the exchange coefficients grouped into symmetry-equivalent shells by interatomic distance. We find that mMACE successfully reproduces the qualitative trends of reference exchange constants reported in~\cite{priessnitz2025ostravaj, rinaldi2024non, rinaldi2023modelling}. This agreement indicates that mMACE captures essential magnetic interactions, despite quantitative differences arising from the approximations used to extract the exchange parameters.

\subsubsection{Curie temperature}
\label{sec:curie_temp}

Iron has an experimental Curie temperature of approximately 1043 K \cite{udovskii2015application}. To assess the predictive capability of different magnetic modeling strategies, we compare Curie temperatures obtained from (i) effective Ising models parameterised by Heisenberg exchange constants $J_s$ (Figs.~\ref{fig:J_total} and \ref{fig:J_adiabatic}) and simulated using \textsc{Vampire}, and (ii) direct finite-temperature simulations with mMACE using spin-flip Monte Carlo (MC) and molecular dynamics–Monte Carlo (MDMC).

The key differences between these approaches are twofold. First, \textsc{Vampire} employs discrete spin-flip updates within a rigid-spin framework, whereas MDMC allows continuous non-collinear spin rotations, enabling the sampling of transverse spin fluctuations. Second, \textsc{Vampire} assumes fixed atomic positions, while MDMC explicitly includes lattice dynamics, thereby accounting for spin–lattice coupling. This treatment more closely resembles finite-temperature \emph{ab initio} approaches \cite{PhysRevLett.121.125902} and allows temperature-dependent effective interactions to emerge naturally.

All models exhibit a collapse of the net magnetisation from approximately 2.2~$\mu_B$/atom to zero near the Curie point (Figs.~\ref{fig:curie_vampire} and \ref{fig:curie_mdmc}). However, Curie temperatures predicted using \textsc{Vampire} vary widely. In particular, models based on adiabatic exchange constants substantially underestimate $T_C$, consistent with the known limitation that infinitesimal spin-rotation approximations typically yield reduced effective exchange interactions.

In contrast, mMACE-driven Monte Carlo (MC) simulations on the non-collinear spin degree of freedom yield Curie temperatures in closer agreement with experimental values. This improvement stems from the explicit treatment of non-collinear spin configurations and higher-order many-body interactions, beyond the two-body exchange terms assumed in the classical Heisenberg model. The resulting temperature--magnetisation curve is benchmarked against the MDMC simulations from the literature performed using ACE. We do not perform MDMC simulations here, as reliable estimation of $T_C$ can practically be sensitive to the MDMC step ratio, which was not reported in the original work. Sampling also the position degrees of freedom requires a much more careful convergence study, which we deferred to future work.

Overall, these results demonstrate that mMACE-based MC simulations provide a more realistic description of finite-temperature magnetism in Fe than conventional rigid-spin Ising-model approaches, capturing essential non-collinear effects.

\section{Discussion and Outlook}
We have introduced mMACE, an equivariant message-passing architecture that extends MACE to incorporate atomic magnetic moments as explicit degrees of freedom. The architecture is equivariant under joint rotations of positions and moments, with spin--orbit coupling effects captured naturally; the stronger $O(3) \times O(3)$ symmetry appropriate for systems without SOC can be enforced through data augmentation or an alternative architecture presented in the appendix. On public benchmarks (FeAl, CrN) mMACE substantially reduces force and stress errors relative to both standard MACE and mMTP baselines. Pre-trained models on the MATPES and MP-ALOE foundational datasets yield systematic improvements on magnetically active configurations while maintaining accuracy elsewhere, and serve as effective starting points for targeted fine-tuning.

The fine-tuning studies on FeNi and \ce{Mn3Pt} illustrate two complementary strengths of the approach. For FeNi, a small number of targeted configurations suffices to recover the Bain-path energetics across ferromagnetic, antiferromagnetic, and non-magnetic states, together with accurate lattice parameters and elastic constants. For \ce{Mn3Pt}, mMACE reliably recovers the frustrated kagome-like non-collinear ground state from random initial spin orientations, demonstrating that the learned energy landscape is qualitatively correct even far from the training distribution. The magnetocrystalline anisotropy study shows that mMACE can resolve SOC-induced energy differences at the sub-meV scale across chemically diverse random structures, achieving relative errors of approximately 5\% and preserving the ranking of anisotropy strengths. Finally, the Fe case study demonstrates that mMACE-driven Monte Carlo simulations yield Curie temperatures in closer agreement with experiment than classical Heisenberg models, benefiting from the explicit treatment of non-collinear configurations and many-body interactions.

Limitations to be addressed in future work include: First, the pre-trained models are fitted to datasets dominated by collinear, predominantly ferromagnetic configurations; extending the foundational training data to include antiferromagnetic, non-collinear, and frustrated magnetic orderings would broaden the applicability of pre-trained mMACE models. Second, while we have demonstrated that SOC-induced anisotropy can be fitted as a perturbative correction, integrating it self-consistently into the training procedure remains an open challenge. Third, extending the model architecture, along the lines of Appendix~\ref{app:sec:non_soc_model}, to effectively treat both non-SOC and SOC systems without the need for data augmentation would benefit both accuracy and generality. 

Finally, coupling mMACE to spin-lattice dynamics frameworks would open the door to studying magnon-phonon interactions, demagnetisation, and other dynamical phenomena beyond the static relaxations and Monte Carlo sampling employed here.

\begin{acknowledgments}
CH, JD, GC, CO and CvdO would like to acknowledge Materials Nexus for providing computational resources with regards to DFT data generation for the random structure search dataset for studying magnetocrystalline anisotropy.
\end{acknowledgments}

\section*{Author contributions}

CH developed the mathematical foundations and code implementation of mMACE and conducted all computational experiments. CH also led the database generation, manuscript writing, and created all figures/tables. CvdO and JD contributed to interpreting the results and guided the project on a daily basis, with CvdO taking a leading role and contributing sections to the manuscript. TK, RB, CE, RK and ES provided suggestions of generating datasets and generated datasets for studying the anisotropy constant. TK and RT validated the mMACE implementation and TK, RT, RB, MP and GC provided project guidance on a weekly basis. RF and JB suggested test cases and systems, and provided project oversight. GC and CO assisted in developing the mathematical foundations of the mMACE architecture. JB, GC and CO provided general project oversight.

\section*{Competing interests}

GC, CO, CvdO, JD are partners of Symmetric Group LLP that licenses force fields commercially. GC also has equity interest in \AA{}ngstr\"om AI.

\clearpage

\appendix

\section{Implementation details}
\label{sec:implementation_details}
\subsection{One-body magnetic contribution in MLIPs}
\label{}\label{app:sec:one_body_mag_contribution_mlips}

The fitting of one-body magnetic contributions can suffer from overfitting or result in non-smooth curves when optimized directly with gradient-based methods such as Adam or LBFGS, which can compromise the physicality of the model. In this appendix, we explain how to obtain a physical $E(m)$ via regularization. To compare more precisely, all plots are shifted by the isolated atom energy at zero magnetic moment so that the one-body magnetic contributions of different elements can be compared. We apply fixed-spin moment (FSM) via occupation control in GPAW to obtain isolated atom reference energies for different magnetic states, enforcing the magnetic moment constraint by adjusting the orbital occupations within the SCF loop and fixing the difference between spin-up and spin-down electron numbers.

\begin{figure}[t]

\centering
    \begin{subfigure}{0.23\textwidth}
        \includegraphics[width=\linewidth]{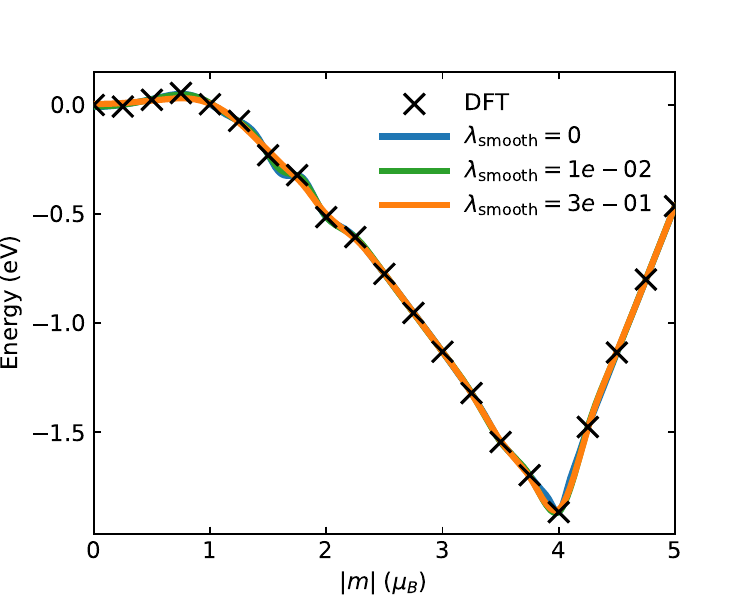}
        \caption{Fe}
        \label{fig:Em_fe}
    \end{subfigure}
    \hfill
    \begin{subfigure}{0.23\textwidth}
        \includegraphics[width=\linewidth]{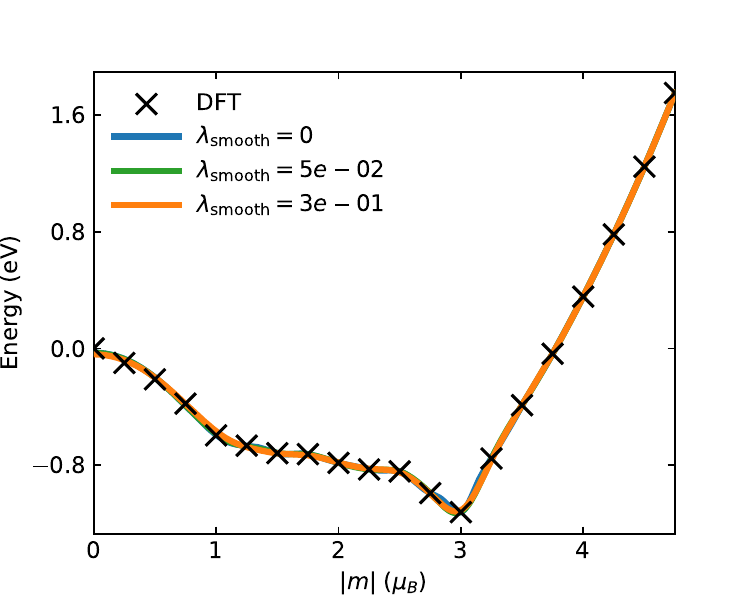}
        \caption{Co}
        \label{fig:Em_co}
    \end{subfigure}

    \vspace{6pt}

    \begin{subfigure}{0.23\textwidth}
        \includegraphics[width=\linewidth]{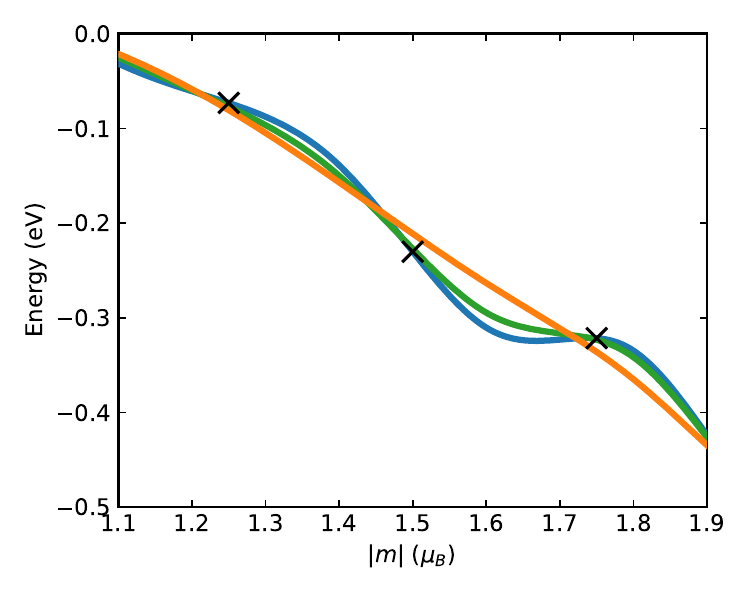}
        \caption{Fe (zoom)}
        \label{fig:Em_fe_zoom}
    \end{subfigure}
    \hfill
    \begin{subfigure}{0.23\textwidth}
        \includegraphics[width=\linewidth]{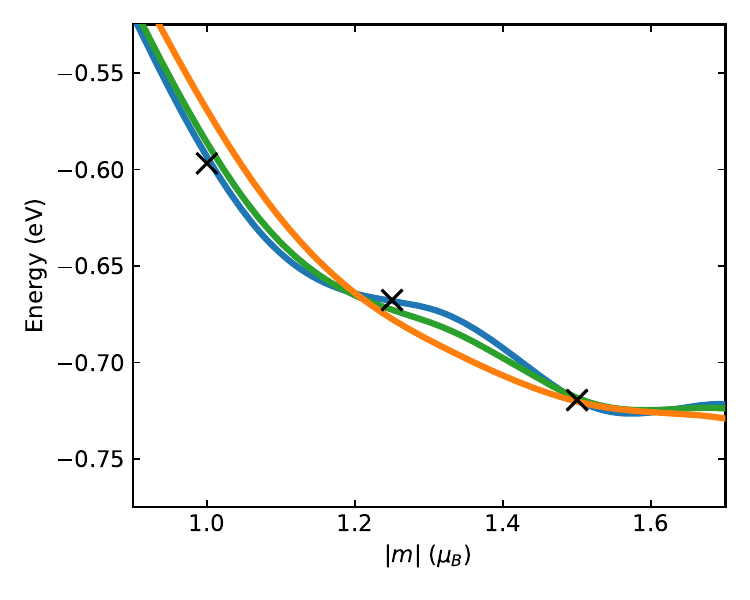}
        \caption{Co (zoom)}
        \label{fig:Em_co_zoom}
    \end{subfigure}

    \vspace{6pt}

    \begin{subfigure}{0.23\textwidth}
        \includegraphics[width=\linewidth]{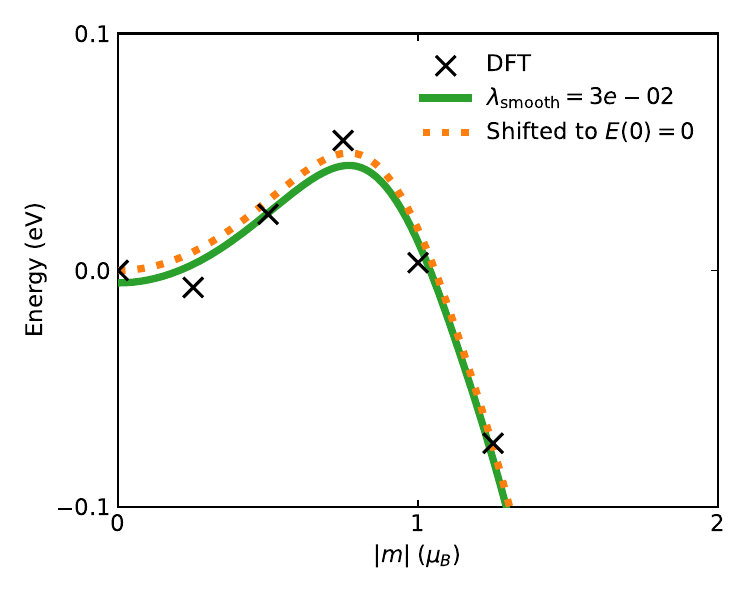}
        \caption{Fe (shifted)}
        \label{fig:Em_fe_shift}
    \end{subfigure}
    \hfill
    \begin{subfigure}{0.23\textwidth}
        \includegraphics[width=\linewidth]{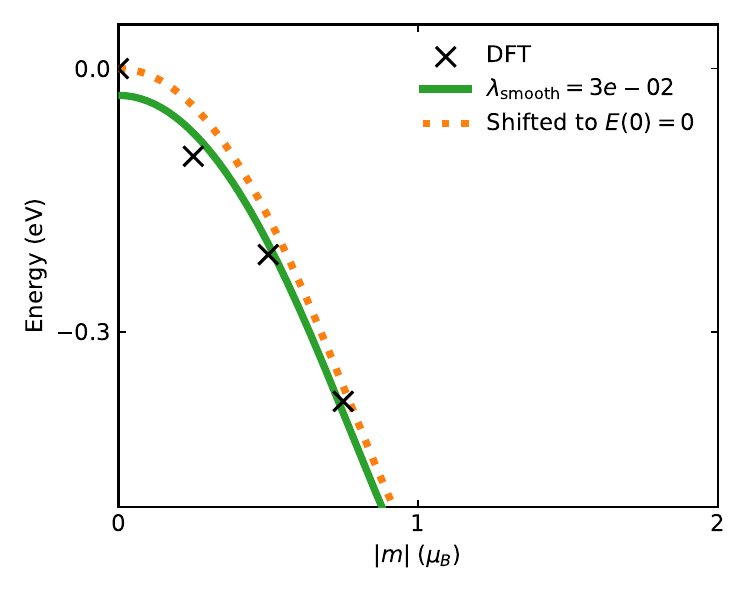}
        \caption{Co (shifted)}
        \label{fig:Em_co_shift}
    \end{subfigure}

    \caption{
    Regularization of one-body magnetic energy curves using curvature-controlled fitting. Panels (a) and (b) show the fitted one-body magnetic energy $E(m)$ for Fe and Co, respectively, obtained from DFT reference data (markers) using different smoothness parameters $\lambda$. Panels (c) and (d) present zoomed-in views of selected regions of panels (a) and (b), highlighting how increasing regularization progressively suppresses short-wavelength curvature while preserving the overall trend of the DFT reference. (e) and (f) demonstrate how the resulting $E(m)$ can be shifted to 0 so that the model reduces to a non-magnetic many-body expansion exactly.
    }
    \label{fig:CO_regularization_scan}
\end{figure}

To obtain a smooth $E(m)$ one-body contribution, one can regularize the coefficients in the linear model $E(m)$ in one of the following ways. Suppose $E(m)$ is simply a Chebyshev polynomial expansion; one can solve the following matrix equation obtained from a regularized least-squares system
\begin{equation}
    (\mathbf{X}^\top \mathbf{X} + \lambda \boldsymbol{\Gamma}^2) \mathbf{c} = \mathbf{X}^\top \mathbf{y},
\end{equation}
where $\mathbf{X} \in \mathbb{R}^{N \times B}$ is the Chebyshev basis evaluated at $N$ sampled magnetic moments with $B$ basis functions, $\boldsymbol{\Gamma} \in \mathbb{R}^{B \times B}$ encodes a regularity prior~\cite{darby2026regularity}, and $\mathbf{y} \in \mathbb{R}^{N}$ contains the DFT-computed one-body magnetic energies. The regularization matrix is diagonal,
\begin{equation}
    \Gamma_{kk} = k^p + 1,
\end{equation}
where $k$ is the polynomial degree and $p$ controls the strength of the smoothness penalty, normally chosen as $p = 4$. For each $\lambda$, the coefficients $\mathbf{c} \in \mathbb{R}^{B}$ are computed via direct linear algebra, and the fitted curves are compared against DFT data to visualize the trade-off between accuracy and smoothness induced by regularization, as demonstrated in Figure~\ref{fig:CO_regularization_scan}.

A more automated way of regularizing $E(m)$ is to augment the loss function, which we found works most robustly when combined with cubic spline interpolation. More precisely, the one-body magnetic energy \(E(m)\) is obtained by fitting a smooth interpolant to the discrete DFT reference points. In the present implementation, this is achieved using an even cubic spline representation \(E(m)=f(m^2)\), which enforces invariance under spin inversion by construction. The fitting objective consists of a data fidelity term combined with a curvature penalty that suppresses large second derivatives of the spline with respect to its argument. Concretely, the optimized spline coefficients minimize
\begin{equation}
    \mathcal{L}_{\rm iso}
    =
    \frac{1}{N}\sum_{i=1}^{N}\bigl(E(m_i)-y_i\bigr)^2
    +
    \lambda \sum_{j}\bigl(\Delta^2 f_j\bigr)^2,
\end{equation}
where $y_i$ denote the DFT reference energies evaluated at sampled magnetic moments $m_i$, and $\Delta^2 f_j$ represents the discrete second finite difference of the spline coefficients, providing a numerical approximation to the curvature of the energy with respect to $m^2$. The regularization parameter \(\lambda\) controls the relative weight of the smoothness penalty: small values allow the spline to closely interpolate the reference data, while larger values progressively suppress oscillations and enforce a more globally smooth energy landscape. By similarly inspecting the fitted curves obtained across a range of \(\lambda\), one can identify a regime in which the resulting \(E(m)\) is both smooth and faithful to the underlying DFT data, providing a robust one-body magnetic reference suitable for subsequent use in the machine-learned potential. The effect of different regularization strengths is demonstrated with two example elements Fe and Co in Figure~\ref{fig:CO_regularization_scan}.

Finally, to match the non-magnetic contribution exactly, the origin of the resulting fit can optionally be shifted to the isolated atom energy at zero magnetic moment, which yields regularized and smoother curves with the correct limit as $|m|\rightarrow0$, as shown in Figures~\ref{fig:Em_fe_shift} and~\ref{fig:Em_co_shift}. The techniques discussed above provide sufficient flexibility for designing the one-body magnetic contribution of MLIPs.

\subsection{Radial basis}
\label{app:sec:radial_basis}

In this section, we give additional details for the construction of the radial bases involved in the above discussion in \eqref{eqn:pos_radial}, \eqref{eqn:mag_radial} and \eqref{eqn:one_body_magmom}. The construction of the Bessel function $J$ is identical to regular, non-magnetic MACE; we refer the reader to \cite{batatia2022mace} for details.

For the magnetic radial basis, let $m_{z_i} \in \mathbb{R}$ be the magnitude of the magnetic moment of an atom with species $z_i$. $T$ maps a single input tuple $(m, z_i)$ to a vector of length equal to the pre-defined maximum degree of the Chebyshev polynomial of the first kind, excluding the constant, via
\begin{equation}
    T(m, z_i) := T(x_{z_i}(m))
\end{equation}
where $x_{z_i}$ is a species dependent transformation
\begin{align}
x_{z_i}(\cdot): m \rightarrow 1 - 2\left( \frac{m}{M_{{\rm max}, z_i}} \right)^2
\end{align}
so that it is mapped from $[0, M_{\max, z_i} ] \rightarrow [-1, 1]$. The constant $M_{\max, z_i}$ is pre-computed depending on the range of magnetic moment magnitudes in the data. When unspecified in the paper, we choose
\begin{equation}
    M_{\max, z_i} = a M^{(data)}_{\max, z_i} + b
\end{equation}
where $a > 1$, $b > 0$ and $M^{(data)}_{\max, z_i}$ is the upper bound of the magnetic moments in the dataset. In the above experiments, we chose $a = 1.2$ and $b = 0.1$, and found that errors and simple in-distribution properties are insensitive to these parameters.

\subsection{Parameter Estimation}
\label{sec:parameter_Estimation}

The term \(E_{0, z_i}(|\mathbf{m}_i|)\) is a one-dimensional function that depends only on the magnitude of the magnetic moment of atom \(i\). Notably, estimating \(E_{0, z_i}\) directly via gradient-based optimization can lead to severe overfitting due to the low intrinsic dimensionality of this term. To mitigate this, we fit the corresponding linear coefficients separately before fitting the remaining part of the model, using regularized approaches, as detailed in Appendix~\ref{app:sec:one_body_mag_contribution_mlips}.

Apart from the two-stage estimation for the one-body magnetic contribution, the remainder of the training procedure follows standard MACE hyperparameters and optimization strategies as described in \cite{batatia2025foundation}. Further details on dataset generation and hyperparameters for parameter estimation can be found in Appendix~\ref{app:sec:parameter_estimation}.

\subsection{Data Augmentation}
\label{app:sec:data_augmentation}
Given that the model is capable of capturing spin–orbit coupling, fitting to datasets without spin–orbit effects (or collinear DFT data) requires the use of data augmentation. In general, we find that data augmentation ensures the preservation of symmetry, introducing deviations of less than 0.1 meV/atom, with only a few rare outliers of comparable magnitude. The artificially introduced SOC discrepancy of the mMACE model is one or two orders of magnitude smaller than the corresponding prediction error. At the same time, the level of symmetry error violation is also in line with \cite{mazitov2025pet}, where data augmentation is similarly applied for training the corresponding MLIPs.
\begin{figure}
    \centering
    \includegraphics[width=0.95\linewidth]{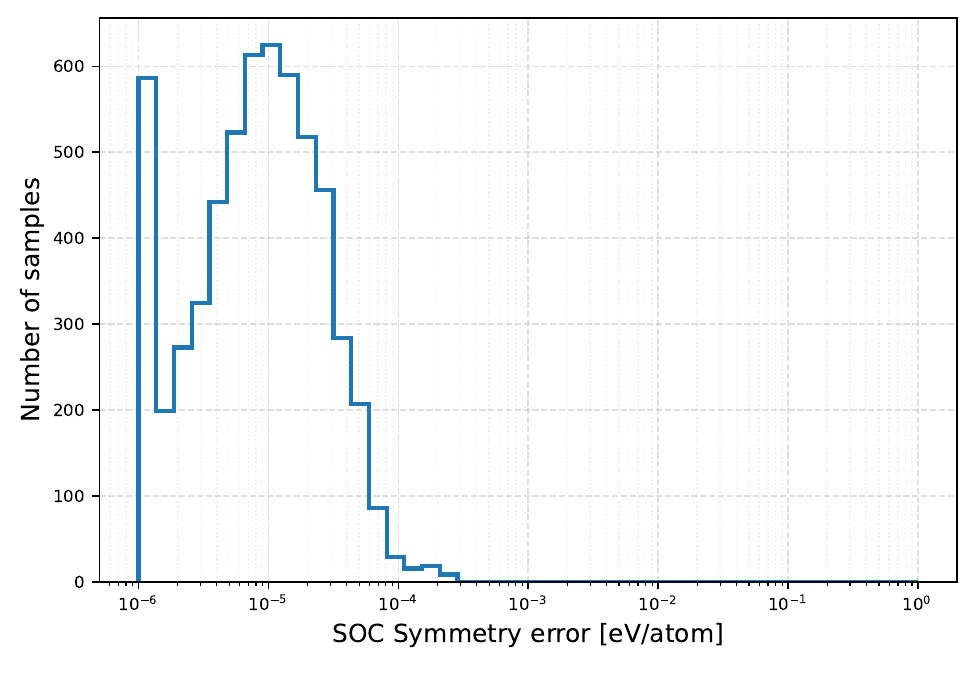}
    \caption{Distribution of the per-atom energy deviation induced by random spatial rotations in mMACE models on the FeNi dataset after fine-tuning. The histogram is shown on a logarithmic scale and quantifies the symmetry error between unrotated and rotated configurations. The vast majority of samples exhibit deviations below 0.1 meV/atom, with only a small number of rare outliers, demonstrating that data augmentation preserves rotational symmetry to high numerical accuracy.}
    \label{fig:symmetry_error_example}
\end{figure}

\section{Magnetic Anisotropy Constant}
\label{app:sec:magnetocrystalline_anisotropy}

The most rigorous approach to computing magnetic anisotropy energy (MAE) within DFT involves performing \textit{non-collinear self-consistent-field} (SCF) calculations with SOC and constrained magnetic moments for each magnetization direction, using fully relativistic pseudopotentials or all-electron methods.
Such calculations are computationally demanding and difficult to converge.

For most materials, however, magnetocrystalline anisotropy (MCA) is effectively a collinear phenomenon: atomic spins remain parallel or antiparallel, and only the orientation of the \textit{net} magnetization changes. Therefore, the SOC contribution to the total energy can often be approximated from \textit{non self-consistent} calculations based on a collinear reference state, where the SOC-induced energy correction is evaluated second-variationally from the scalar relativistic eigenstates.

A common and computationally efficient approach for evaluating such energy differences is the \textit{magnetic force theorem} (MFT). Within the MFT approximation, the change in total energy due to SOC or a rotation of the magnetization direction can be estimated from the difference in non-self-consistent band energies obtained using the eigenstates of a collinear calculation \cite{mortensen2024gpaw}:
\begin{equation}
\label{eqn:mft_mae}
    {\rm MAE} \approx 
    \sum_i f_i(\theta_1, \phi_1)\, \epsilon_i(\theta_1, \phi_1)
    - \sum_i f_i(\theta_0, \phi_0)\, \epsilon_i(\theta_0, \phi_0),
\end{equation}
where $\epsilon_i(\theta, \phi)$ and $f_i(\theta, \phi)$ denote the eigenvalues and occupation numbers of the Kohn–Sham states for magnetization direction $(\theta, \phi)$, respectively. This method exploits the variational property of the total energy with respect to the density, allowing one to bypass the expensive SCF procedure for each spin orientation.

Within the mMACE framework, the same orientation-dependent band-energy difference is learned directly from reference SOC DFT data from MFT. Typically, the scale of SOC effect is sub-meV per atom (see Fig.~\ref{fig:soc_distribution}), which is comparable to or smaller than the typical errors of modern MLIPs in single-point energy predictions. Nevertheless, within the standard workflow based on the magnetic force theorem, the task of approximating $E_{A}$ reduces to estimating the SOC-induced difference in band energy between a magnetization orientation $(\theta, \phi)$ and a reference orientation $(\theta_0, \phi_0)$.
\begin{align}
\label{eqn:diff_band_energy_main}
E_{A}(\theta, \phi; \{\sigma_i\}_i) & \approx \delta E_{\rm band, soc} (\theta, \phi; \{\sigma_i\}_i) \nonumber \\
&= E_{\rm band, soc}(\theta, \phi; \{\sigma_i\}_i) \nonumber\\
& - E_{\rm band, soc}(\theta_0, \phi_0; \{\sigma_i\}_i).
\end{align}
where $\{\sigma_i\}_i :=  \{ (\mathbf{r}_i, \mathbf{m}_i, z_i)\}_i$ denotes a single configuration.

This allows the model to approximate the SOC-induced anisotropy landscape in a fully differentiable and computationally efficient manner, without explicitly invoking relativistic electronic-structure calculations.

\section{Heisenberg exchange constants}
\subsection{Total energy difference method for Heisenberg exchange constant}
\label{app:sec:heisenberg_exchange_constant}

In what follows we describe how Heisenberg exchange constants can be obtained from a trained mMACE model. We consider the Heisenberg Hamiltonian of the system 
\begin{equation}
    \mathcal{H} := \mathcal{H}_{0} - \sum_{\langle ij \rangle} J'_{ij} \hat{\mathbf{S}}_{i} \cdot \hat{\mathbf{S}}_{j}
\end{equation}
where $\mathcal{H}_0$ denotes the non-magnetic part of the energy and $J'_{ij}$ is the exchange interaction between spins $\tilde{S}_{i}$ and $\tilde{S}_{j}$. The summation goes over all spin pairs once. By considering symmetry and equivalent crystal sites (with equivalence classes)
\begin{align}
    \mathcal{H}_{\rm Heis} 
    & = \sum_{k} J_k c_k \\
    c_k & = \sum_{\langle ij \rangle \in X_k} \hat{\mathbf{S}}_{i} \cdot \hat{\mathbf{S}}_{j}
\end{align}
In practice, a finite cutoff $d_{\rm max}$ is chosen, which results in a finite number of equivalence classes $N_k$ and all other exchange interaction energies are considered zero. The Heisenberg Hamiltonian then becomes
\begin{equation}
    \mathcal{H}_{\rm Heis} = \sum_{k < N_k} J_k c_k
\end{equation}
For each magnetic configuration (indexed by $l$), one can construct an expression for the energy $E^{l}$ with $J_k$ as unknown
\begin{equation}
    E^{(l)} = E_0 - \sum_{k < N_k} J_k c^{(l)}_k,
\end{equation}
where $E_0$ is the non-magnetic part of the total energy. This leads to a linear least squares system from which we can estimate $J_k$, taking the following form
\begin{equation}
    \mathbf{E} = E_0 \mathbf{1} - \mathbf{C}\mathbf{J},
\end{equation}
where $\mathbf{E} = [E^{(1)}, E^{(2)}, \ldots, E^{(L)}]^{\mathrm{T}}$ contains the total energies from $L$ magnetic configurations and $\mathbf{C}_{lk} = c^{(l)}_k$ are the corresponding spin–spin correlation coefficients obtained from the dot products $\hat{\mathbf{S}}_i \cdot \hat{\mathbf{S}}_j$ summed over each neighbor shell $k$.

In practice, we include $E_0$ directly as an additional parameter to be fitted, rather than eliminating it by mean subtraction. This is achieved by appending a column of ones to the design matrix, resulting in the augmented linear system
\begin{equation}
    \mathbf{E} = \mathbf{C}' \mathbf{x},
    \qquad
    \mathbf{C}' =
    \begin{bmatrix}
        -c^{(1)}_1 & \cdots & -c^{(1)}_{N_k} & 1 \\
        \vdots & \ddots & \vdots & \vdots \\
        -c^{(L)}_1 & \cdots & -c^{(L)}_{N_k} & 1
    \end{bmatrix},
\end{equation}
where $\mathbf{x} =(J_1 \cdots J_{N_k} E_0)$. The parameters are then determined by solving the above least–squares problem, which yields both the exchange constants $\{J_k\}_{k < N_k}$ and the reference energy $E_0$ simultaneously.

\subsection{Adiabatic approximation for Heisenberg exchange constant}
\label{sec:app:Adiabatic_approximation_for_Heisenberg_exchange_constant}

In the real-space approach of Liechtenstein et al.~\cite{liechtenstein1984exchange, liechtenstein1986lsdf, liechtenstein1987local}, the adiabatic magnon energy of a spin spiral with wavevector
$\mathbf{q}$ is given by
\begin{equation}
    E(\mathbf{q}) = \sum_{j} J_{0j}\,\Bigl[1 - \cos\!\bigl(\mathbf{q}\cdot \mathbf{r}_{0j}\bigr)\Bigr],
    \label{eq:adiabatic_magnon_energy}
\end{equation}
where $J_{0j}$ are the real–space Heisenberg exchange interactions between a reference atom and neighbours at displacement $\mathbf{r}_{0j}$.

To extract individual $J_{ij}$ values, we use the small–angle rotation protocol. Starting from a collinear ferromagnetic
state, the magnetic moments of sites $i$ and $j$ are rotated by $\pm \theta/2$ in opposite directions,
while all other spins remain aligned. Denoting the resulting total–energy variations by
\[
    \delta E_i(\theta), \qquad
    \delta E_j(\theta), \qquad
    \delta E_{ij}(\theta),
\]
the pair interaction energy is defined as
\begin{equation}
    \Delta_{ij}(\theta)
    = \delta E_{ij}(\theta)
      - \bigl[\delta E_i(\theta) + \delta E_j(\theta)\bigr].
    \label{eq:delta_ij_def}
\end{equation}
For sufficiently small rotation angles, the Heisenberg model predicts the quadratic relation
\begin{equation}
    \Delta_{ij}(\theta)
    = \tfrac{1}{2} J_{ij}\, \theta^2 + \mathcal{O}(\theta^4).
    \label{eq:small_angle_fit}
\end{equation}
Hence, $J_{ij}$ can be obtained via a linear fit of $\Delta_{ij}(\theta)$ versus $\theta^2/2$.
This provides a computationally efficient route to estimating exchange interactions without
explicitly constructing long–wavelength spirals. 

It is important to emphasize that the effective Heisenberg exchange constants obtained from these calculations are not necessarily identical to those derived from the total energy difference method. Each approach serves only as an estimation of the exchange interaction through a different physical pathway, and the resulting values may therefore differ systematically, as reflected by the numerical experiments in Section~\ref{app:sec:heisenberg_exchange_constant}.

\section{Experiment supplements}

\subsection{CrN public benchmark}
\label{app:sec:CrN_mMACE_magforce_hist2d}

\begin{figure}
    \centering
    \hspace{-0.5cm}
    \includegraphics[width=0.95\linewidth]{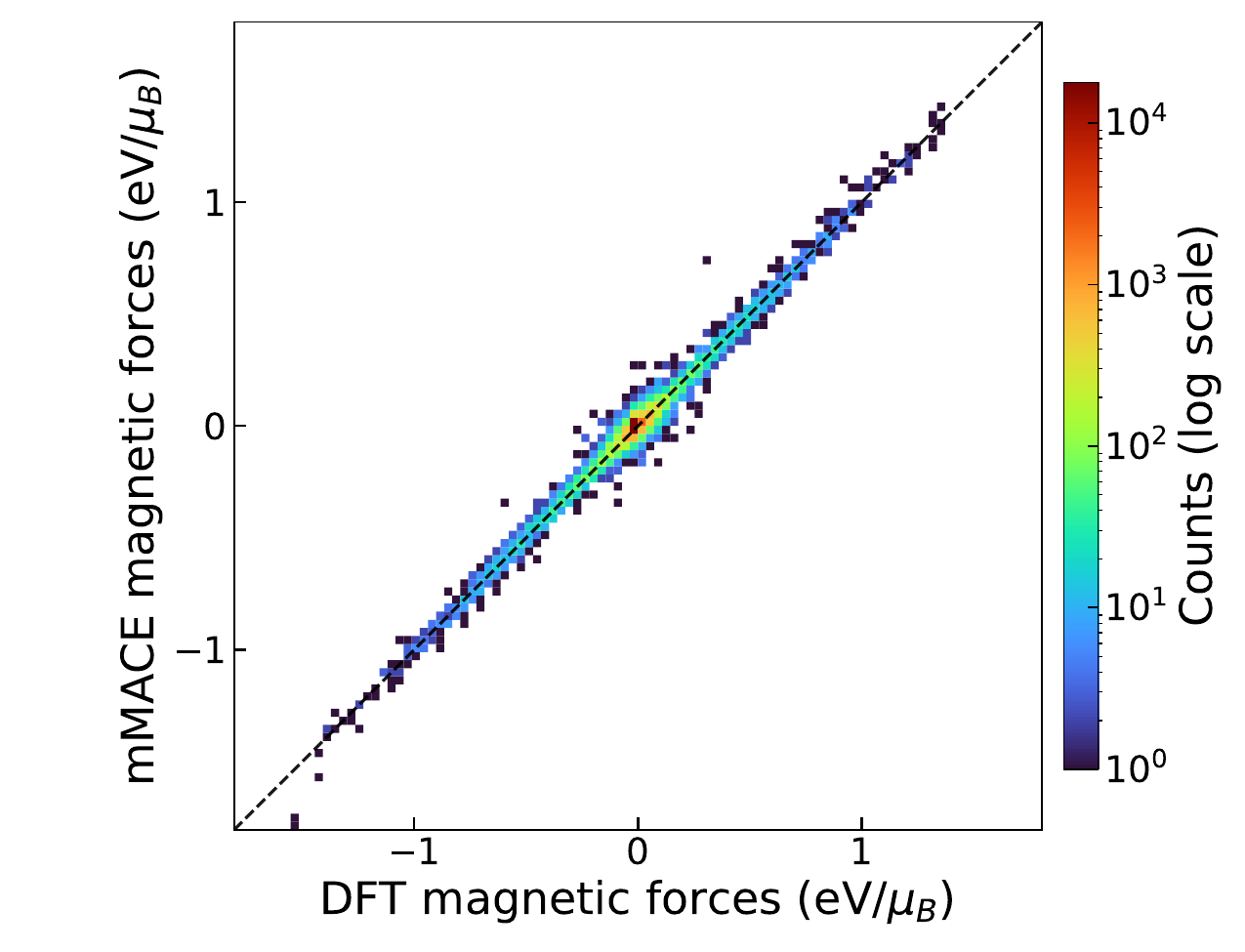}
    \caption{2D histogram comparing mMACE to DFT reference magnetic forces on the CrN dataset.}
    \label{fig:CrN_mMACE_magforce_hist2d}
\end{figure}

Figure~\ref{fig:CrN_mMACE_magforce_hist2d} shows a strong concentration of points along the diagonal, indicating excellent agreement between mMACE-predicted and DFT magnetic forces over the full force range. The narrow spread around the identity line, even in the low-force regime, demonstrates that mMACE accurately captures both the magnitude and sign of magnetic force responses in CrN.

\subsection{Energy minimization on pre-trained datasets}
\label{app:relaxed_magnetic_moments}

Single-point evaluation metrics do not fully capture the behavior of magnetic degrees of freedom under minimization of the magnetic moments. The ground-state energy defined for an atomic configuration should be minimized with respect to the underlying internal variable $\{ \mathbf{m}_i\}_i$ representing local magnetic moment. Different initial local magnetic moments can lead to qualitatively different relaxed spin configurations, especially in systems with competing magnetic states.

To quantify this effect, we analyze the consistency between energy-minmized and reference collinear magnetic moments across the MATPES-PBE, MATPES-r2SCAN, and MP-ALOE datasets. A magnetic moment minimization (i.e. relaxation) is classified as \emph{fail} if the resulting minimization gives \emph{NaN}, the energy goes above 100 eV / atom or forces go above 30 eV/\AA. Around $2.02\%$ (MATPES-PBE), $1.72\%$ (MATPES-r2SCAN), and $1.29\%$ (MP-ALOE) of the energy minimizations fail on the testing set. This is expected because we are only sampling a relatively narrow range of magnetic configurations from a small dataset; we hypothesize that including constrained DFT and/or applied fields would greatly increase the magnetic diversity.

\begin{table}[h]
\centering
\begin{tabular}{lccc}
\toprule
Dataset & Energy & Force & Stress \\
\midrule
MATPES-PBE    & 55.6 & 130.8 & 0.9969 \\
MATPES-r2SCAN & 61.8 & 154.4 & 1.1515 \\
MP-ALOE       & 70.8 & 100.8 & 1.4239 \\
\bottomrule
\end{tabular}
\caption{Error metrics on testing set across the MATPES-PBE, MATPES-r2SCAN, and MP-ALOE datasets, reported as mean absolute errors (MAE) (units: energy in meV/atom, forces in meV/\AA, stress in GPa).}
\label{tab:sp_metrics}
\end{table}

\begin{figure*}[t]
    \centering
    \begin{subfigure}[t]{0.32\textwidth}
        \centering
        \includegraphics[width=\linewidth]{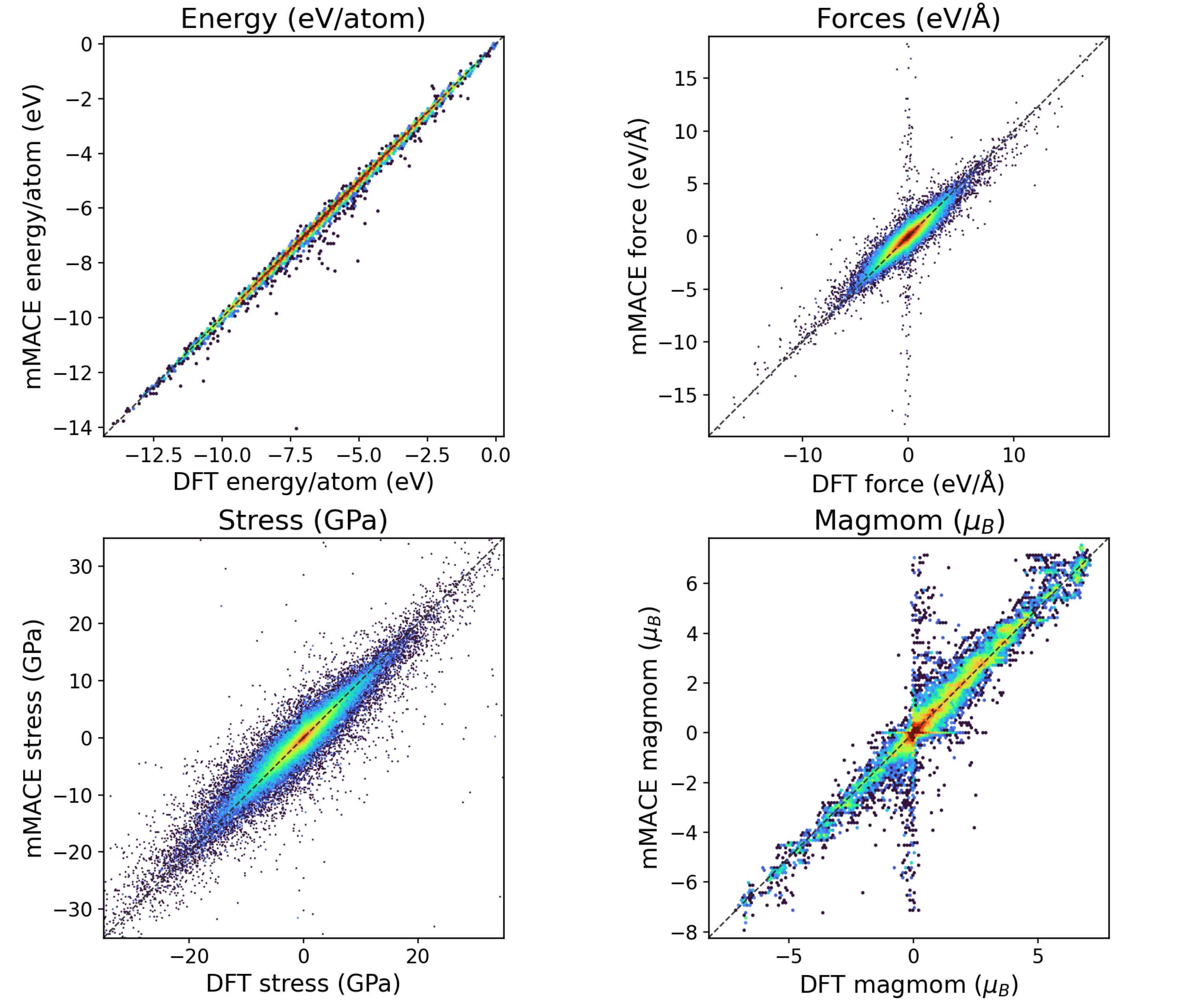}
        \caption{MATPES-PBE}
    \end{subfigure}
    \hfill
    \begin{subfigure}[t]{0.32\textwidth}
        \centering
        \includegraphics[width=\linewidth]{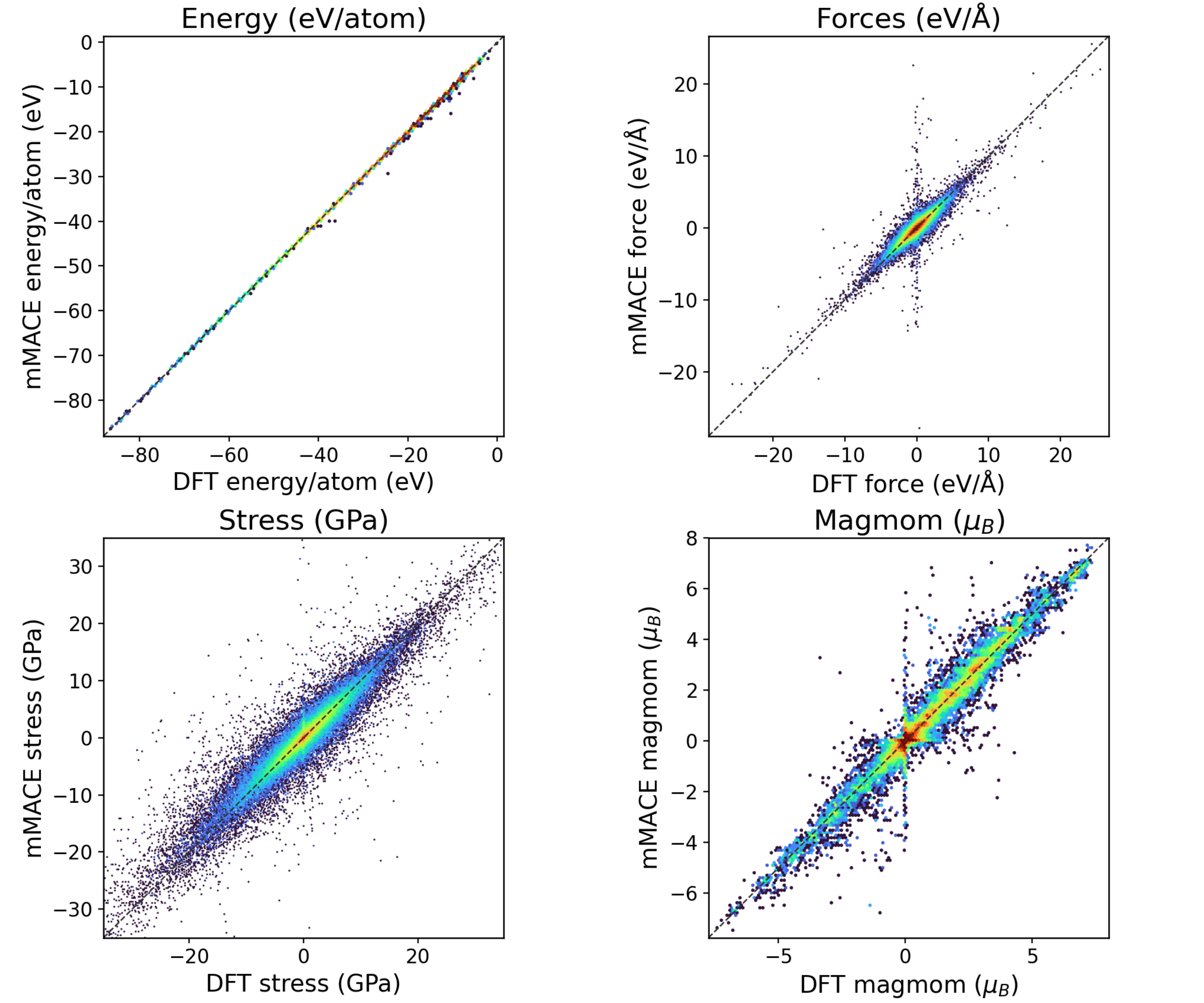}
        \caption{MATPES-r2SCAN}
    \end{subfigure}
    \hfill
    \begin{subfigure}[t]{0.32\textwidth}
        \centering
        \includegraphics[width=\linewidth]{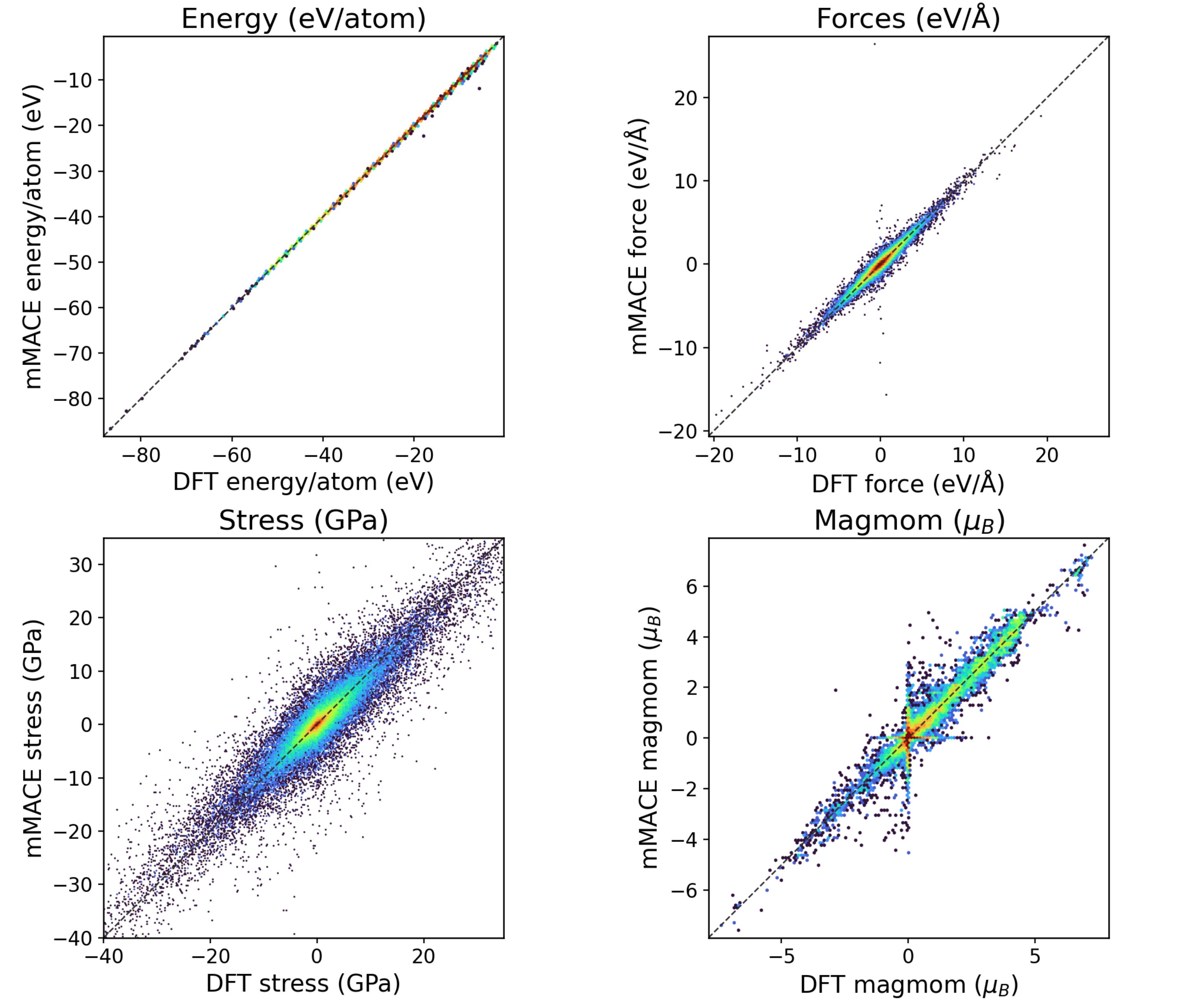}
        \caption{MP-ALOE}
    \end{subfigure}
    \caption{
    Parity plots for energy, forces, stress, and magnetic moments across the three datasets. Energies and forces show strong agreement with DFT, while magnetic moments exhibit increased dispersion, particularly near $m \approx 0$, reflecting the presence of competing magnetic states.
    }
    \label{fig:parity_full}
\end{figure*}

For completeness, the corresponding accuracy on the subsets where the minimizations were successful are summarized in Table~\ref{tab:sp_metrics}. Figure~\ref{fig:parity_full} shows parity plots for energy, forces, stress, and magnetic moments across the three datasets. While energies and stresses exhibit strong agreement with DFT, the magnetic moments display a broader distribution, particularly in the at around $m_\mathrm{DFT} \approx 0$. 

These discrepancies are not merely due to inaccurate model, but reflect an intrinsic ambiguity in the mapping between atomic structure and equilibrium magnetic configuration within spin DFT, leading to inconsistent dataset. More precisely, the mapping
\begin{equation}
    \mathbf{R} \mapsto \{\mathbf{m}^*_i\}
\end{equation}
where $\mathbf{m}^*$ indicates a set of local magnetic moment from self-consistent magnetization density, is not single-valued.

We also note that local magnetic moments are not uniquely defined quantities, but depend on the partitioning scheme used to decompose the electronic density. Different schemes (e.g., Wigner-Seitz, Hirshfeld, or Bader-type approaches) can assign different values to atomic moments even when derived from the same charge density. This dependence is particularly relevant in systems with delocalized magnetization or strong hybridization, where the spatial distribution of spin density is diffuse.

\paragraph{Representative failure case.}
We illustrate this behavior using a representative 3-atom TbSbPt bulk configuration from the MATPES-PBE test set (MATPES id: \texttt{mp-16313}). The reference DFT solution corresponds to a nearly non-magnetic state with total moment $m \approx 0.34\,\mu_B$, whereas mMACE predicts a strongly spin-polarized configuration with a local moment of approximately $5.67\,\mu_B$ on the Tb site, indicating a qualitative change in magnetic state.

To determine which solution is energetically favorable, due to DFT software availability, we perform self-consistent DFT calculations in GPAW (rather than VASP used in MATPES) using two different initializations: (i) the reference DFT magnetic moments and (ii) the mMACE-predicted magnetic configuration. The resulting total energies are $E_{\mathrm{DFT}}^{\mathrm{ref}} = -16.55~\mathrm{eV}$ and $E_{\mathrm{DFT}}^{\mathrm{mMACE}} = -21.84~\mathrm{eV}$, showing that the spin-polarized state obtained from the mMACE initialization is significantly lower in energy.

\begin{table}[h]
\centering
\begin{tabular}{lccc}
\toprule
Method & Tb & Sb & Pt \\
\midrule
DFT (VASP reference)            & 0.339   & 0.001     & 0.001     \\
mMACE (predicted)          & 5.67    & 0.0004    & 0.363     \\
GPAW (init from DFT)       & -0.002  & -0.00017  & -0.0012   \\
GPAW (init from mMACE)     & 5.82    & -0.0098   & 0.0013    \\
\bottomrule
\end{tabular}
\caption{Converged local magnetic moments (in $\mu_B$) for TbSbPt under different initializations.}
\label{tab:magmom_tb_sb_pt}
\end{table}

Table~\ref{tab:magmom_tb_sb_pt} shows that the two initializations converge to distinct self-consistent solutions: a non-magnetic state when initialized from the DFT reference, and a strongly spin-polarized state when initialized from the mMACE prediction. The close agreement between the mMACE-predicted moments and the DFT solution obtained from the same initialization further indicates that mMACE identifies a stable magnetic branch of the electronic structure, rather than producing an unphysical configuration.

\paragraph{Internal inconsistency, role of similar structures and implications for MLIPs.}
To understand why mMACE converges to a strongly spin-polarized state for the TbSbPt configuration, we examine whether similar magnetic environments are present in the MATPES training data. We find that the dataset contains structurally similar configurations with magnetic states closely matching the mMACE prediction.

In particular, structurally similar bulk configurations that is derived from the same entry in Materials Project as the above structure exhibits a spin states (with mean absolute error up to $0.2 \mu B$.) that is very close to mMACE-relaxed value. The existence of these configuration in the training set, together with similar bulk configuration such as TbNiSb (MATPES id: \texttt{mp-3716}) that also exhibits a similar local magnetic moment, contributes to the fact that mMACE converges to the lower energy spin states with lower DFT energy of $-21.57$ eV in the TbSbPt structure as given by GPAW in Table~\ref{tab:magmom_tb_sb_pt} above.


More broadly, this highlights a fundamental limitation when fitting non-magnetic MLIPs to datasets such as MATPES: they implicitly mix multiple self-consistent magnetic solutions of the Born-Oppenheimer surface without distinguishing between them. As a result, similar atomic environments can correspond to qualitatively different magnetic states and substantially different energies, violating the assumption of a single-valued mapping $E(\mathbf{R})$.

From a modeling perspective, this introduces intrinsic ambiguity into the learning problem. Apparent errors in predicted magnetic moments may instead reflect selection of an alternative valid magnetic branch rather than model failure. In this case, the mMACE prediction corresponds to a state that is both self-consistently stable under DFT and lower in energy than the provided reference.

These observations suggest that magnetic datasets should be interpreted as sampling a multi-branch energy landscape unless only the global minimum branch across different configurations is included, and that evaluation of magnetic MLIPs should go beyond single-point comparisons to incorporate relaxation-based validation or branch-aware metrics.

\subsection{FeNi lattice parameters}
\label{apd:sec:feni_lattice_parameters}
As shown in Table~\ref{tab:mmace_ref_lattice_mag_elastic}, equilibrium lattice parameters agree with DFT within 0.01~\AA\ across all phases and magnetic states. Elastic constants are reproduced with correct trends and typical deviations of ~10\%, with larger errors confined to non-magnetic phases due to limited sampling.

\begin{table*}[ht!]
\centering
\renewcommand{\arraystretch}{1.2}
\setlength{\tabcolsep}{6pt}
\begin{tabular}{llccccc}
\hline
\textbf{Structure} & \textbf{Mag.}
& $a$ (\AA)
& $C_{11}$
& $C_{12}$
& $C_{44}$
& $\epsilon_{M_z}$ ($\mu_B$/atom) \\
\hline

fcc & FM  
& 3.57 (3.57)
& 1.666 (1.507)
& 1.080 (0.885)
& 0.812 (0.787)
& 0.007 \\

fcc & AFM 
& 3.57 (3.56)
& 1.637 (1.332)
& 1.071 (0.961)
& 0.815 (0.644)
& 0.141 \\

fcc & NM  
& 3.50 (3.50)
& 2.230 (2.228)
& 1.280 (1.259)
& 0.964 (1.181)
& 0.000 \\

bcc & FM  
& 3.59 (3.59)
& 1.992 (2.026)
& 0.273 (0.243)
& 0.865 (0.789)
& 0.101 \\

bcc & AFM 
& 3.59 (3.58)
& 1.992 (1.628)
& 0.272 (0.189)
& 0.868 (0.776)
& 0.013 \\

bcc & NM  
& 3.51 (3.51)
& 1.833 (2.176)
& 0.408 (0.421)
& 0.892 (1.121)
& 0.000 \\

\hline
\end{tabular}
\caption{
Equilibrium lattice constants $a$, magnetization error $\epsilon_{M_z}$,
and elastic constants $C_{11}$, $C_{12}$, and $C_{44}$ (in GPa)
for bcc and fcc FeNi under different magnetic states.
Elastic constants are reported as mMACE predictions with reference (DFT) values shown in parentheses.
}
\label{tab:mmace_ref_lattice_mag_elastic}
\end{table*}

\subsection{\ce{Mn3Pt} relaxations}
\label{app:sec:Mn3Pt_relaxations}
To perform the relaxation experiment, we first set up a fixed atomic configuration from a reference geometry optimization of the standard cubic Mn$_3$Pt structure. We then apply random 3D rotations to the reference magnetic moments with angles sampled uniformly from $[0, \pi]$, followed by uniform random scaling from $[0, 1.5]$ on each Mn atom. Exactly 100 such magnetic orientations are generated, on which we perform relaxation of the magnetic moments with the LBFGS optimizer until the maximum gradient reaches $10^{-4}$. The magnetic moment of the Pt atom is kept fixed as it is known to be highly non-magnetic in this system, with a local magnetic moment on the order of $10^{-4}\,\mu_B$.

\subsection{MDMC simulation details}
\label{app:sec:conv_mdmc}

\begin{figure}[t]
    \centering

    \begin{subfigure}{0.98\linewidth}
        \centering
        \includegraphics[width=\linewidth]{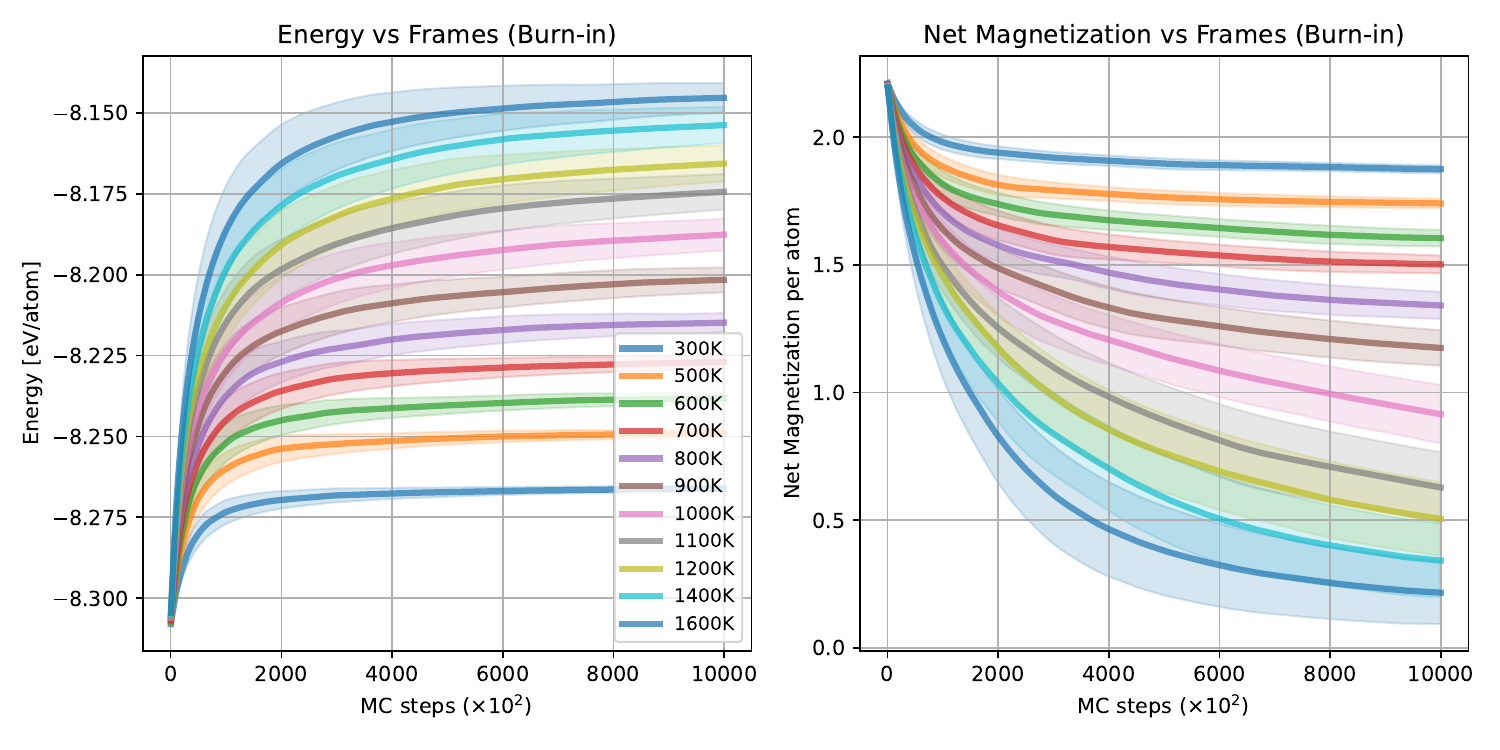}
        \caption{Burn-in stage to equilibrate the system.}
        \label{fig:mdmc_stage1}
    \end{subfigure}

    \vspace{1em}

    \begin{subfigure}{0.98\linewidth}
        \centering
        \includegraphics[width=\linewidth]{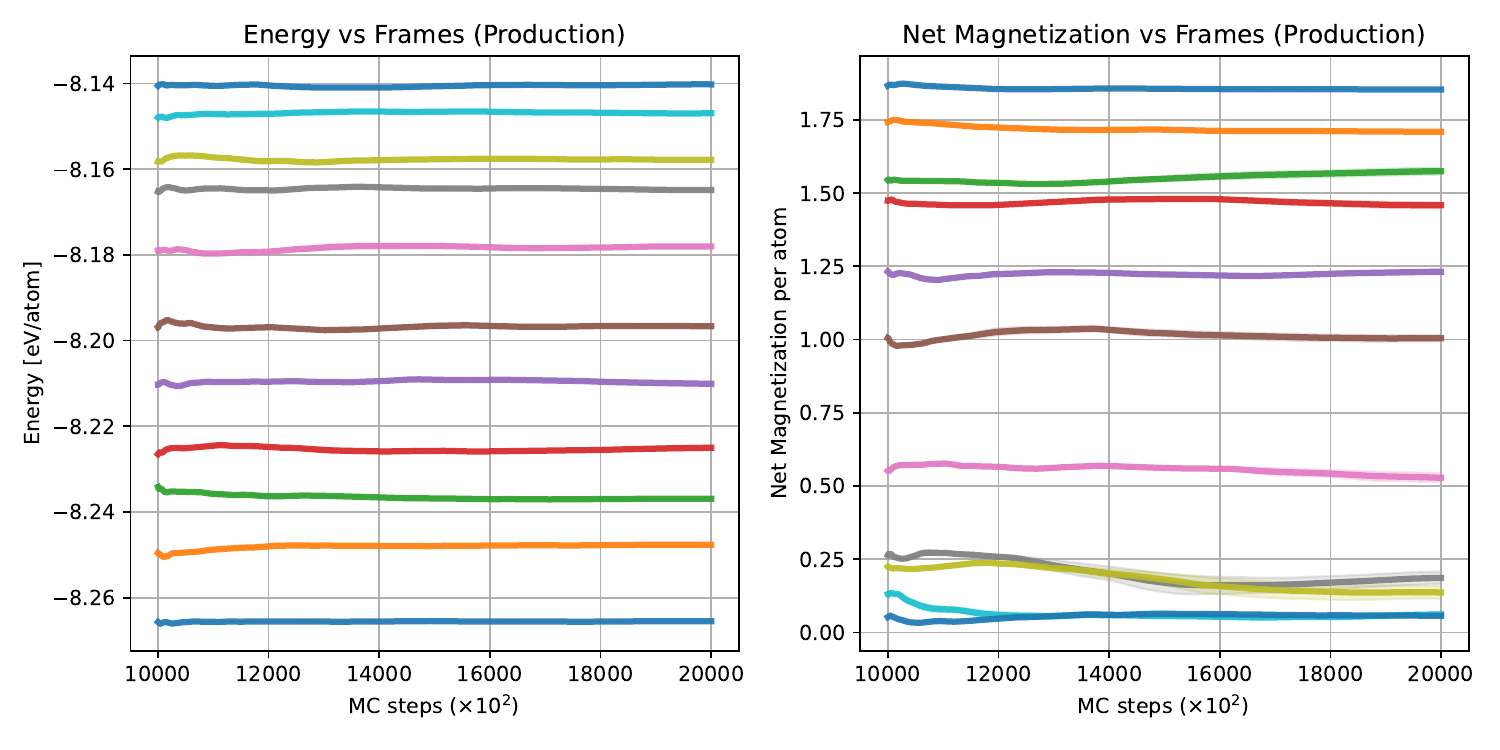}
        \caption{Production stage for statistical averaging.}
        \label{fig:mdmc_stage2}
    \end{subfigure}

    \caption{Convergence behavior of MC simulations.}
    \label{fig:mdmc_convergence}
\end{figure}

Figure~\ref{fig:mdmc_convergence} demonstrates convergence of the MC simulation in production and burn-in stages as mentioned in Section~\ref{sec:curie_temp}.

Curie temperature calculations were performed using a bcc Fe supercell containing 3456 atoms. The Curie temperature $T_C$ was extracted by fitting $(T - T_C)^{\beta}$ where both $T_C$ and the critical exponent $\beta$ were treated as fitting parameters. 

The Curie temperature was estimated by fitting $(T - T_C)^{\beta}$ where both $T_C$ and $\beta$ were treated as fitting parameters. The \textsc{Vampire} simulations were performed using $J_{ij}$ for up to 6 and 5 nearest neighbours for the adiabatic and total-energy methods, respectively, to match the number of Heisenberg exchange constants included in existing references. We found that the results tend to be insensitive to exchange interactions beyond the 4th neighbour.

\section{A Non-SOC mMACE Variant}
\label{app:sec:non_soc_model}
We briefly summarize a variant of mMACE that implements the more restrictive non-SOC symmetry \eqref{eq:nospinorbitcoup}. 
The main modification is that the tensors' $(l, m)$ channels are replaced by $(l, m, l', m')$ channels to independently encode rotations acting on positions and magnetic moments. 
With notation as in Section~\ref{sec:theory}, for the first layer ($t = 1$) we initialize the features as a one-hot embedding of the chemical species
\begin{equation}
    h^{(1)}_{j, k0000} := \sum_{z} W_{k z_j}\delta_{z z_j}.
\end{equation}
This defines the node state $\tilde{\sigma}_i := (\mathbf{r}_i, \mathbf{m}_i, z_i, h^{(t)}_i)$ for message passing, as in \eqref{eqn:general_node_state}. 

With identical initialization of $P^{(t)}_{ji, k l_1 l_2 l_3}$ and $K^{(t)}_{ji, k l_1 l_2 l_3}$ as in \eqref{eqn:pos_radial} and \eqref{eqn:mag_radial} respectively, the two-body features with magnetic information can be constructed through the following sequence of operations:
\begin{align}
\phi^{(t),{\rm pos}}_{ji, k l_3 m_3}
&:= \sum_{l_1 m_1 l_2 m_2} 
    C^{l_3 m_3}_{l_1 m_1 l_2 m_2}  \\
    \notag
&\hspace{1.6cm} \cdot P^{(t)}_{ji, k l_1 l_2 l_3}  Y^{m_1}_{l_1}(\hat{\mathbf{r}}_{ji}) 
    h^{(t)}_{j, k l_2 m_2 0 0} \\
\phi^{(t),{\rm mag}}_{ji, k l'_3 m'_3}
&:= \sum_{l_1' m_1' l_2' m_2'} 
    C^{l'_3 m'_3}_{l'_1 m_1' l'_2 m_2'}  \\
    \notag
&\hspace{1.6cm} \cdot K^{(t)}_{ji, k l'_1 l'_2 l'_3}  R^{m'_1}_{l'_1}(\hat{\mathbf{r}}_{ji}) 
    h^{(t)}_{j, k 00 l'_2 m_2'} \\
\phi^{(t),{\rm mag+pos}}_{ji, k l m l' m'} 
&:= \phi^{(t),{\rm pos}}_{ji, k l m} \phi^{(t),{\rm mag}}_{ji, k l' m'} \\
A^{(t)}_{i, k  l m l' m'}
&:= \sum_{j \in \mathcal{N}(i)} 
    \phi^{(t),{\rm mag + pos}}_{ji, k l m l' m'}.
\end{align}
Here, $\hat{\mathbf{r}}_{ji} := \mathbf{r}_{ji} / |\mathbf{r}|$, $Y^{m_1}_{l_1}$ are real spherical harmonics, and $R^{m'_1}_{l'_1}$ are solid harmonics which guarantee smoothness in the limit $|\mathbf{m}| \rightarrow 0$. 

Once the two-body basis $A^{(t)}_{i, klml'm'}$ is formed, the many-body basis for each correlation order is constructed as
\begin{align}
\mathbf{B}^{(t)}_{i, \eta_\nu \eta'_\nu k L M L' M'} & := \sum_{\mathbf{l'm'}} C_{\eta'_\nu \mathbf{l'm'}}^{L' M'} \sum_{\mathbf{lm}} C_{\eta_\nu \mathbf{lm}}^{L M} \notag \\
& \prod_{\xi=1}^{\nu} \sum_{\tilde{k}} 
W^{(t)}_{k \tilde{k} l_{\xi} l_{\xi}'} A_{i,\tilde{k}  l_\xi m_\xi l'_\xi m'_\xi}^{(t)},
\end{align}
where, $C^{L M}_{\eta_\nu \mathbf{l m}}$ are generalized Clebsch–Gordan coefficients~\cite{batatia2023general}. Note that in this case, no coupling is involved between $\mathbf{l}$ and $\mathbf{l}'$, hence the CG contractions can be performed independently and parallelized. The many-body message is obtained by summing over contributions from each correlation order $\nu$
\begin{equation}
    \mu^{(t)}_{i,k L M L' M'} := \sum_{\nu} \sum_{\eta_\nu \eta'_\nu} W_{z_i \eta_\nu \eta'_\nu k L L'}^{(t), \nu} \mathbf{B}^{(t)}_{i, k\eta_\nu \eta'_\nu L M L' M'}.
\end{equation}
To include the contribution from the magnetic moment of atom $i$ itself and correlate it with its atomic environment, an additional contraction is performed, 
\begin{align}
    &\tilde{\mu}^{(t)}_{i, k l m L' M'} \\ 
    \notag
    &:= \sum_{l_0 m_0 l' m'} C^{L' M'}_{l_0 m_0 l' m'} T^{(t)}_{i, k l_0 l' L'}R^{m_0}_{l_0}(\mathbf{m}_i) \mu^{(t)}_{i, k l m l' m'}.
\end{align}
Here $T^{(t)}_{i, k l_0 l' L}$ is again a learnable radial function analogous to \eqref{eqn:mag_radial} but depends only on the species and magnetic moment magnitude of atom $i$.

Across layers, the node feature update is completed via a residual connection, following standard practice in modern neural network architectures
\begin{align}
    \tilde{h}^{(t+1)}_{i, k LM L' M'}
    := & \sum_{\tilde{k}} W_{kLL',\tilde{k}}^{(t),\mathrm{self}} \tilde{\mu}_{i,\tilde{k}LM L' M'}^{(t)} \notag \\
    & + \sum_{\tilde{k}} W_{kLL',\tilde{k}}^{(t)} \mu_{i,\tilde{k} LM L' M'}^{(t)} \notag \\
    & + \sum_{\tilde{k}} W_{k z_i L L',\tilde{k}}^{(t)} h_{i,\tilde{k}LM L' M'}^{(t)}.
\end{align}
The total energy is then obtained as a sum of readouts (either linear or non-linear MLPs) applied to the invariant node features. Additionally, a one-body magnetic moment contribution is introduced to ensure the correct large volume limit (\textit{i.e.} as $|\mathbf{r}_{ji}| \rightarrow \infty$ for all $i,j$). This term is pre-trained or fitted as a simple linear expansion of Chebyshev polynomials as discussed in the main model architecture as described in Section~\ref{sec:theory}:
\begin{align}
\label{eqn:one_body_magmom_app}
    \notag 
    E_i 
    &:= \sum_{t} \left( \mathcal{R}_t\left(h^{(t)}_{i, \bar{k}LM L'M'}\right) \right) + \sum_{n z} W_{n z} \delta_{z z_i} T_n(x(|\mathbf{m}_i|)) \\
    & =:  E^{\rm inter}_i + E_{0, z_i}(|\mathbf{m}_i|).
\end{align}
This completes a single forward pass of a non-SOC model.

\section{Hyperparameters for parameter estimation}
\label{app:sec:parameter_estimation}
Unless otherwise stated, all models are optimized with Adam optimizer with learning rate $0.005$. AMSgrad and EMA are applied with default parameters in the MACE package. A gradient clip of 100 is applied. Models are trained using a composite loss function that balances energies, forces, and stresses, with relative weights $w_E = 1$, $w_F = 10$, and $w_\sigma = 5$, respectively. Data augmentation is applied whenever the underlying DFT of the input data does not include SOC effects. A list of major hyperparameters for the model architecture is summarized in Table~\ref{table:mMACE_major_hyperparameters}.

\begin{table*}[ht]
\centering
\begin{tabular}{ll}
\toprule
\textbf{Hyperparameter} & \textbf{Description} \\
\midrule
Hidden irreps & Feature dimension in the latent representation \\
Number of layers & Depth of the message-passing architecture \\
Body order ($\nu$) & Maximum order of many-body correlations included \\
Cutoff radius ($r_{\max}$) & Spatial range of atomic interactions \\
Angular order (position, $\ell^{\rm pos}_{\max}$) & Maximum angular resolution for positional features \\
Angular order (magnetic, $\ell^{\rm mag}_{\max}$) & Maximum angular resolution for magnetic features \\
Radial basis size (position) & Number of radial basis functions for positions \\
Radial basis size (magnetic) & Number of radial basis functions for magnetic moments \\
Magnetic cutoff ($M_{\max, Z}$) & Maximum allowed magnetic moment for atoms of species type $Z$ \\
\bottomrule
\end{tabular}
\caption{Summary of key hyperparameters in the mMACE architecture.}
\label{table:mMACE_major_hyperparameters}
\end{table*}

\subsection{FeAl public dataset}
We employ a single layer model with scalar hidden irreducible representation \texttt{64$\times$0e}. Body order correlations are included up to order $\nu = 3$, with a real-space interaction cutoff of $r_{\max} = 5.0~\text{\AA}$. Angular dependencies of position and magnetic moments are resolved using spherical harmonics up to $\ell^{\rm pos}_{\max} = 3$ and $\ell^{\rm mag}_{\max} = 1$ respectively. The radial dependence of position and magnetic degree of freedom is expanded in a 10 and 8 Bessel basis and Chebyshev basis respectively. The magnetic moments are restricted to the range $M_{\max, {\rm Al}}=0.15$ and $M_{\max, {\rm Fe}}= 4.0$ respectively. We also enforce one-body contributions to be simply constant for fair benchmarking with mMTP results without additional data.

\subsection{CrN public dataset}
We employ a single layer model with scalar hidden irreducible representation \texttt{64$\times$0e}. Body order correlations are included up to order $\nu = 3$, with a real-space interaction cutoff of $r_{\max} = 5.0~\text{\AA}$. Angular dependencies of position and magnetic moments are resolved using spherical harmonics up to $\ell^{\rm pos}_{\max} = 3$ and $\ell^{\rm mag}_{\max} = 1$ respectively. The radial dependence of position and magnetic degree of freedom is expanded in a 10 and 8 Bessel basis and Chebyshev basis respectively. The magnetic moments are restricted to the range $M_{\max, {\rm Al}}=0.15$ and $M_{\max, {\rm Fe}}= 4.0$ respectively. We also enforce one-body contributions to be simply constant for fair benchmarking with mMTP results without additional data.

\subsection{Pretrained models on foundational datasets}
We employ a two layer model with scalar hidden irreducible representation \texttt{128$\times$0e + 128 $\times$1o}. Body order correlations are included up to order $\nu = 3$, with a real-space interaction cutoff of $r_{\max} = 6.0~\text{\AA}$. Angular dependencies of position and magnetic moments are resolved using spherical harmonics up to $\ell^{\rm pos}_{\max} = 3$ and $\ell^{\rm mag}_{\max} = 1$ respectively. The radial dependence of position and magnetic degree of freedom is expanded in a 10 and 8 Bessel basis and Chebyshev basis respectively. The magnetic moments are restricted to the range $M_{\max, {\rm Al}}=0.15$ and $M_{\max, {\rm Fe}}= 4.0$ respectively. We also enforce one-body contributions to be simply constant for fair benchmarking without additional data to existing MACE architecture.

\subsection{FeNi dataset}
Identical model hyperparameters to the pre-trained model MATPES-PBE are used as the model is fine-tuned from the MATPES-PBE mMACE model. A lower learning rate of 0.0005 and a gradient clip of 10 instead of 100 are applied.

\subsection{Magnetocrystalline anisotropy datasets}
For fitting both FeMn and the general database constructed from random structure search, we employ a two layer model with scalar hidden irreducible representation \texttt{64$\times$0e + 64$\times$1o}. Body order correlations are included up to order $\nu = 3$, with a real-space interaction cutoff of $r_{\max} = 6.0~\text{\AA}$. Angular dependencies of position and magnetic moments are resolved using spherical harmonics up to $\ell^{\rm pos}_{\max} = 3$ and $\ell^{\rm mag}_{\max} = 3$ respectively. The radial dependence of position and magnetic degree of freedom is expanded in a 10 and 8 Bessel basis and Chebyshev basis respectively.

The SOC configuration identification task uses the same model architecture as the MATPES pre-trained model without the constant energy shift.

During training, the validation error is observed to increase monotonically after the initial epoch, indicating rapid overfitting due to the limited number of parent configurations. As a result, we use the model parameters from epoch~0 (i.e., the pretrained MATPES model without further fine-tuning) for all SOC identification results. Despite the absence of additional training, the model already exhibits strong rank correlation with DFT reference SOC energy spreads, demonstrating that the pretrained representation encodes sufficient information to identify SOC-sensitive configurations.

\subsection{\ce{Mn3Pt} dataset}
Identical model hyperparameters to the pre-trained model MATPES-PBE are used as the model is fine-tuned from the MATPES-PBE mMACE model. A lower learning rate of 0.0005 and a gradient clip of 10 instead of 100 are applied.

\subsection{Fe dataset}
We employ a single layer model with scalar hidden irreducible representation \texttt{64$\times$0e}. Body order correlations are included up to order $\nu = 3$, with a real-space interaction cutoff of $r_{\max} = 4.5~\text{\AA}$. Angular dependencies of position and magnetic moments are resolved using spherical harmonics up to $\ell^{\rm pos}_{\max} = 3$ and $\ell^{\rm mag}_{\max} = 3$ respectively. The radial dependence of position and magnetic degree of freedom is embedded in a 10 and 8 Bessel basis and Chebyshev basis respectively before feeding into MLPs. The magnetic moments are restricted to $M_{\max, {\rm Fe}}= 4.0$. 

\FloatBarrier

\section{Data generation}
\label{app:sec:data_generation}

\subsection{Pretrained models on foundational datasets}
\label{app:sec:data_generation:foundation}
A filtering process was performed to filter configurations that contain extreme DFT ground truth forces, as we found it improves the stability of the model in each corresponding reference dataset in \cite{kaplan2025foundational} and \cite{kuner2025mp} respectively. Configurations with forces larger than $20~\text{eV/\AA}$ are removed from the dataset. For MATPES-PBE and MATPES-r2SCAN, training, validation and testing set from the official split with the above filtering. For MP-ALOE, only the configuration from the final ionic step of each configuration are included as those are the only configurations given with magnetic moment.

\subsection{\ce{FeNi} dataset}
\label{app:sec:data_generation:FeNi}

First, we generate a broad family of unit-cell–distorted structures by applying random perturbations to an ordered bcc \ce{FeNi} alloy unit cell, with Fe and Ni arranged in an Fe–Ni–Fe–Ni sequence. The fcc base structures and a structure with cell ratio $a/c = 1.6$ are used to generate 5 variants by applying an isotropic volume scaling with a random scale factor uniformly sampled from $[0.6, 1.4]$, followed by adding Gaussian random displacements with $\sigma = 0.05 \AA$ to all atomic positions. We then assign part of the data an initial guess of magnetic moments $[2.2, 0.6, 2.2, 0.6]~\mu_\mathrm{B}$, and the corresponding AFM magnetic arrangement, corresponding to the equilibrium values for bcc Fe and Ni, respectively. We also apply spin-flip operations, as well as scaling the magnetic moment of each atom with uniform random scaling sampled from $[0, 1]$. This stage provides diverse and reasonably off-equilibrium configurations.

Second, we perform direct DFT relaxations of both base structures and sample intermediate configurations along each relaxation trajectory. These data capture the energetics and magnetic responses along the physically relevant transformation pathway between the two phases and their respective minima. 

This completes the first iteration of dataset generation. We use the resulting dataset to perform a the first round of fine-tuning. The resulting Bain path is shown by the blue line in Figure~\ref{fig:FeNi_BP_all} (labeled as iter1). The poor performance, in particular the model’s confusion between the FM and AFM states, indicates that more extensive sampling is required. 

In the second iteration of data generation we enrich the dataset with off-equilibrium yet physically sensible configurations. To that end, we generate additional structures by applying Monte Carlo–like perturbations to both atomic positions and lattice degrees of freedom. Starting from a reference structure, random Gaussian displacements are applied to the atomic positions, together with perturbations to the simulation cell. Configurations with trial Monte Carlo moves (with admissible moves sampled from random position and cell rattling) are evaluated using the fine-tuned mMACE model, and acceptance is determined via a Metropolis criterion based on the Boltzmann probability at $2000$~K to sufficiently generate a wide range of cell and positions samples. This ensures dense coverage of the $a/c$ ratio and cell volumes relevant to the Bain path, as shown in Figure~\ref{fig:FeNi_BP_all}.

To further improve the quality of the Bain path, we additionally sample the AFM Bain path generated by the previous model. These configurations are subsequently perturbed in atomic positions and magnetic moments, and evaluated using single-point DFT calculations. This final stage enriches the dataset with configurations that are energetically accessible but rarely sampled by direct \textit{ab initio} relaxations.

A second fine-tuning procedure is performed using the resulting dataset, which contains approximately 244 configurations. The resulting Bain path is shown in Figure~\ref{fig:FeNi_BP_all} (labeled as iter2).

Testing set are generated from randomly drawing samples from the configurations generated in iteration 1. For all the calculations in this example, spin-polarized density functional theory calculations were performed using GPAW with a plane-wave basis set and the PBE exchange–correlation functional. A kinetic energy cutoff of $600$ eV was used for the plane-wave expansion. Brillouin-zone integrations were carried out using Monkhorst–Pack grids generated from a reciprocal-space density of approximately $0.03$ \AA$^{-1}$. All calculations were performed within the collinear spin approximation without spin–orbit coupling. Atomic positions were relaxed until the residual forces were below $1 \times 10^{-4}$~eV/\AA, while the simulation cell was kept fixed. Electronic self-consistency was achieved using GPAW’s default convergence settings.
 
\subsection{Magnetic anisotropy constant}
\label{app:sec:data_generation_magnetic_anitostropy}
When generating data in this section, for each dataset all angles are defined with a fixed coordinate frame, i.e., $\phi$ and $\theta$ are always pointing to $x$ and $z$ respectively as it has to be consistent over different parent configurations for the fitting to make sense.

In the experiment that follows, for each parent configuration, we uniformly sample spin orientations over the unit sphere and evaluate the corresponding band energy differences \eqref{eqn:diff_band_energy_main} for each orientation. These energy differences are then used as the training targets, and the fitting procedure is carried out in the same manner as for energy fitting in MLIPs with mMACE.

\subsubsection{\ce{FeMn} dataset}
\label{app:sec:FeMn_soc_datagen}
To construct the training dataset, we start from a fully relaxed collinear DFT structure. From this fixed atomic configuration, we uniformly sample 10,000 magnetization orientations over the unit sphere and compute the corresponding SOC band energies. Calculations are run with spin-polarized DFT in GPAW with the PBE functional with identical settings as \ce{FeNi}. Magnetocrystalline anisotropy energies were computed using \texttt{spinorbit.soc\_eigenstates} by evaluating SOC-induced band energy differences for fixed magnetization directions. We emphasize that this dataset is intentionally over-sampled and is designed primarily to assess the expressiveness and fidelity of the proposed framework, rather than to represent a minimal or production-level training protocol.

\subsubsection{Random structure search (RSS) dataset}
\label{app:sec:RSS_soc_datagen}
We generated a large database of random ternary structures within the chemical space spanned by Al, Mn, Fe, Co, Ni, Y, Pt, optionally combined with one light element from B, C, Si, using random structure search (RSS) with unit cells containing up to 10 atoms. Structure generation was performed using \texttt{buildcell} driven by MACE-MPA-0, and the resulting configurations were screened using MACE-MatPES-PBE-0 to evaluate energies, forces, and pressures.

From this pool, 1000 structures were sampled with a bias to match the elemental distribution of the parent dataset. Magnetic orderings were generated for each structure using the \texttt{MagneticStructureEnumerator} implemented in \texttt{pymatgen}~\cite{Ong2013pymatgen}, version 2025.10.7. These structures were relaxed using DFT (Quantum ESPRESSO~\cite{giannozzi2009quantum}), comprising non-magnetic, ferromagnetic, ferrimagnetic, and antiferromagnetic states.

Magnetic anisotropy calculations were subsequently performed using GPAW~\cite{mortensen2024gpaw}. For nominally non-magnetic systems, the self-consistent calculations were initialized with small random magnetic moments to enable SOC-induced anisotropy. In total, around 242 collinear calculations were used in the subsequent analysis and 2000 are sampled on the sphere for each parent configuration.

\subsection{\ce{Mn3Pt} dataset}
\label{app:sec:Mn3Pt_datagen}
Training structures were generated from the four-atom primitive unit cell of \ce{Mn3Pt}, comprising three Mn atoms forming the frustrated kagome sublattice and one Pt atom occupying the face-centered position. We first performed a reference density-functional theory (DFT) calculation by fully relaxing the structure initialized according to Ref.~\cite{zuniga2023observation}, using Quantum ESPRESSO~\cite{giannozzi2009quantum} with the PBE exchange–correlation functional and non-relativistic pseudopotentials. Single-point density functional theory (DFT) calculations were performed for a total of 147 configurations using Quantum ESPRESSO~\cite{giannozzi2009quantum}, with a plane-wave energy cutoff of 816~eV and a \emph{k}-point density of 0.29~\AA$^{-1}$.

\section{Performance}
\label{app:sec:performance}
In this section, we perform a performance analysis with the current implementation of mMACE against the original MACE, using the MATPES pre-trained model in Section~\ref{sec:pretrained_model_training}. 

\begin{figure}[t]
    \centering
    \begin{subfigure}{0.23\textwidth}
        \centering
        \includegraphics[width=\linewidth]{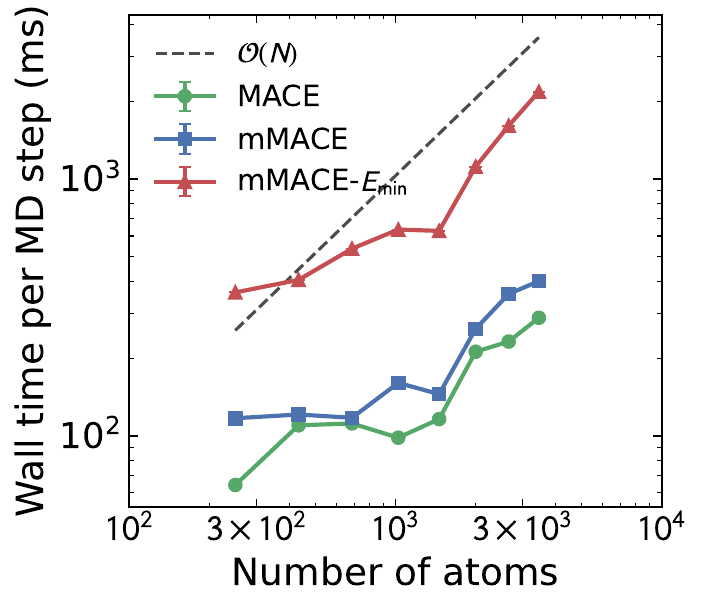}
        \caption{}
        \label{fig:performance_cost}
    \end{subfigure}
    \hfill
    \begin{subfigure}{0.23\textwidth}
        \centering
        \includegraphics[width=\linewidth]{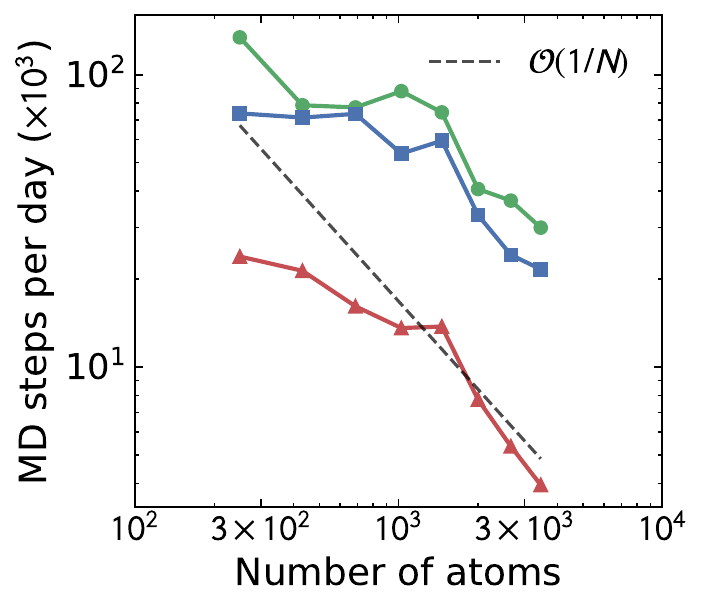}
        \caption{}
        \label{fig:performance_throughput}
    \end{subfigure}
    \caption{
    \textbf{Performance scaling of magnetic interatomic potentials on bcc Fe.}
    (\textbf{a}) Wall-clock time per molecular dynamics step (energy and forces) as a function of system size.
    (\textbf{b}) Corresponding molecular dynamics throughput, reported as the number of MD steps per day.
    Both panels compare the baseline MACE model and its magnetic extension (mMACE), demonstrating near-linear scaling with system size and a moderate overhead associated with the magnetic degrees of freedom.
    }
    \label{fig:performance_scaling}
\end{figure}

To evaluate the difference in performance between original and mMACE we compare the time per MD steps in Fig.~\ref{fig:performance_scaling}. Usual MACE, mMACE and mMACE with energy minimized at every MD step until the gradient of energy with respect to the magnetic moment converges up to $10^{-3}$ are compared. In the case where energy is minimized, the energy between subsequent minimization updates is less than $10^{-3}$ meV / atom. We found that mMACE takes longer to converge in the first energy minimization (for example, on a 250 atoms supercell, it requires 18 energy minimization steps), but requires fewer iteration later (4-10 energy minimization steps), analogous to re-using electron densities in DFT. 

\bibliographystyle{plain}
\bibliography{reference.bib}

\end{document}